\documentclass[preprint2]{aastex}

\newcommand{\myemail}{fadda@ipac.caltech.edu}

\slugcomment{}

\shorttitle{24$\mu$m Spitzer First Look Survey}
\shortauthors{Fadda et al.}

\begin{document}


\title{The Spitzer Space Telescope Extra-Galactic First Look Survey:
24$\mu$m data reduction, catalog, and source identification}


\author{Dario Fadda~\altaffilmark{1}, 
Francine R. Marleau, 
Lisa J. Storrie-Lombardi,\\
David Makovoz, 
David T. Frayer, 
P.N. Appleton,
L. Armus,
S. C. Chapman~\altaffilmark{2},
P. I. Choi,
F. Fang,
I. Heinrichsen,
G. Helou,
M. Im~\altaffilmark{3},
M. Lacy,
D. L. Shupe,
B.T. Soifer,
G. K. Squires,
J. Surace,
H. I. Teplitz,
G. Wilson,
L. Yan
}
\affil{Spitzer Science Center, California Institute of Technology, MC 220-6, Pasadena, CA 91125}

\altaffiltext{1}{\myemail}
\altaffiltext{2}{Department of Astronomy, California Institute of Technology, MS 320-47, Pasadena, CA 91125, USA}
\altaffiltext{3}{Astronomy Program, School of Earth and Environmental Sciences, Seoul National University, Shillim-dong, Kwanak-gu, Seoul,
S. Korea 2-880-9010}



\begin{abstract}

We present the reduction of the 24$\mu$m data obtained during the
first cosmological survey performed by the Spitzer Space Telescope
(First Look Survey, FLS).  Images of a region of sky at moderately
high galactic latitude (l=88.3, b=+34.9) were obtained on 9-11
December 2003. The survey consists of a shallow observation of
$2.5^o\times2^o$ centered at $17^h18^m +59^o30'$ (main survey) and a
deeper observation of $1^o\times0.5^o$ centered at $17^h17^m +59^o45'$
(verification survey).  Issues with the reduction of the 24$\mu$m MIPS data 
are discussed and solutions to attenuate instrumental effects are proposed
and applied to the data.  Approximately 17000 sources are extracted
with a signal-to-noise ratio (SNR) greater than five. The photometry of the
point sources is evaluated through PSF fitting using an empirical PSF
derived from the data.  Aperture corrections and the absolute calibration
have been checked using stars in the field.  Astrometric and
photometric errors depend on the SNR of the source varying between
0.35~--~1 arcsec and 5\%~--~15\%, respectively, for sources detected
at 20~--~5~$\sigma$.The
flux of the 123 extended sources have been estimated through aperture
photometry. The extended sources cover less than 0.3\% of the total
area of the survey.
 Based on simulations, the main and verification surveys
are 50\% complete at 0.3 and 0.15~mJy, respectively.  Counterparts have
been searched for in optical and radio catalogs.  More than 80\% of the
24~$\mu$m sources have a reliable optical counterpart down to
$R=25.5$. 16\% of the sources have a 20~cm counterpart down to 0.1~mJy
and $\sim$80\% of the radio-infrared associations have a reliable
optical counterpart.  A residual map is obtained by subtracting point
sources detected at the 3$\sigma$ level and interpolating the regions
occupied by extended sources. Several galactic clouds with low and
intermediate velocities are identified by comparison with neutral
Hydrogen data from this field.

\end{abstract}


\keywords{catalogs --- surveys ---  infrared:galaxies}


\section{Introduction}

On December 2003, during its first scientific campaign, the Spitzer
Space Telescope (Werner et al., 2004) completed its first
extragalactic survey in the far-IR. Almost 27 hours of observations
with MIPS (Multiband Imaging Photometer for Spitzer, Rieke et al. 2004)
were used to scan several times 5 square degrees of sky in the
direction 17$^h18^m$ +59$^o30'$.  The
region was observed simultaneously in three different bands (24$\mu$m, 70$\mu$m and
160$\mu$m), thanks to the multiband capability of MIPS.

In this paper, we present the reduction and analysis of the 24$\mu$m
data. Although the far-IR bands sample better the maximum of emission
of the IR sources, the 24$\mu$m band is the best band for
extragalactic studies thanks to the high sensitivity of the 24$\mu$m
array and to the fair spatial resolution (5.9'' PSF FWHM) of the
instrument.  IRAC (InfraRed Array Camera, Fazio et al. 2004)
data were also obtained. These have been utilized  to help in the identification of optical
counterparts and to better define the spectral energy distribution of
the detected sources (Lacy et al., 2005).

\begin{figure}[h!]
\setlength{\fboxsep}{3.5cm}
\framebox[8cm]{See figure 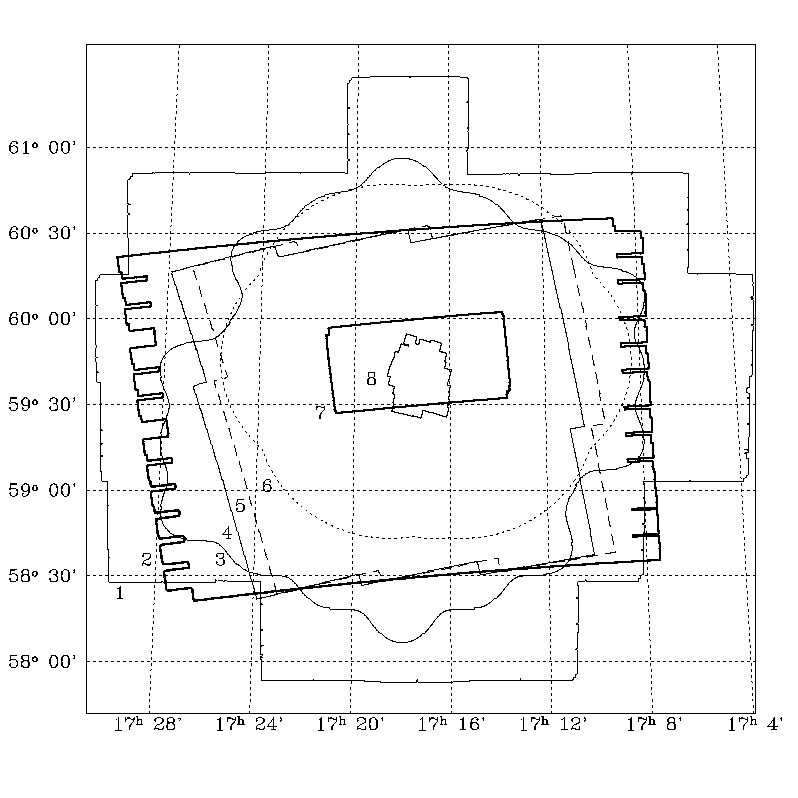}
\caption{Coverage of public surveys in the FLS region. From the
outside going toward the center: (1) KPNO R-images (Fadda et al. 2004), (2)
24 $\mu$m main field (this paper), (3) VLA 20cm (Condon et al., 2003),
(4) IRAC channels 1 and 3 (Lacy et al. 2005), (5) IRAC channels 2 and
4 (Lacy et al. 2005), (6) WSRT 20cm (Morganti et al. 2004), (7) 24
$\mu$m verification field (this paper), (8) ACS/HST data
(Storrie-Lombardi et al., in prep.).  A region similar to the 24$\mu$m
one has been covered also at 70$\mu$m and 160 $\mu$m (Frayer et
al. 2005).
\label{fig:surveys}
}
\end{figure}

The improvement of depth and quality of the mid-IR images with Spitzer
is huge with respect to previous space missions. IRAS (Soifer et al. 1983)  surveyed
the same region of sky at 25$\mu$m detecting only 6 sources.  Although
ISO (Kessler et al.\ 1996) did not observe the FLS field, several
surveys have been conducted at 15$\mu$m. The most extended in the
ELAIS-S1 and Lockman regions (Gruppioni et al.\ 2002, Fadda et
al.\ 2004) covered 2 and 0.55 sq.deg., respectively. They reached a
depth of 1 and 0.45 mJy detecting 100 and 480 sources per square degree (31 and
5 times less than the source density in the main field and 76 and 16 times less than the source density in the verification field), respectively.  Only the
deepest surveys (like the HDF-North, Serjeant et al. 1997) reached a
sensitivity comparable to that of the First Look Survey but on an area 500
times smaller than that of the FLS verification field.  Moreover, the
different wavelength coverage makes possible the detection of galaxies
up to redshift of 2.5, a region unaccessible to the ISO 15$\mu$m
surveys.

This substantial leap in observational capabilities led to the execution 
of the First Look Survey using Directors Discretionary Time. The
early release of data from this survey gave 
general observers real data to assist in planning of their 
Cycle-1 proposals rather than basing extragalactic observations 
on highly extrapolated models.  A subset
of the data was released in January 2004 and the full 
pipeline processed data set was released when the Spitzer archive opened
(May 2004).  This paper presents the state-of-art reduction based on
tools developed at the Spitzer Science Center (SSC).  

The main goals of the surveys were: (i) to detect enough extragalactic
sources at unexplored sensitivity levels in order to generate a
representative sample and to reduce the uncertainties in source counts
(see Marleau et al. 2004); (ii) characterize the dominant source
populations with both MIPS and IRAC data from Spitzer, plus ancillary
data at optical, near-infrared and radio wavelengths (see e.g. Yan et
al., Appleton et al. 2004) and (iii) explore the cirrus foreground at
moderately high galactic latitudes, and its effect on point-source
detectability.

\begin{figure*}[t!]
\hbox{
\includegraphics[angle=0,width=8cm,clip=true,angle=0]{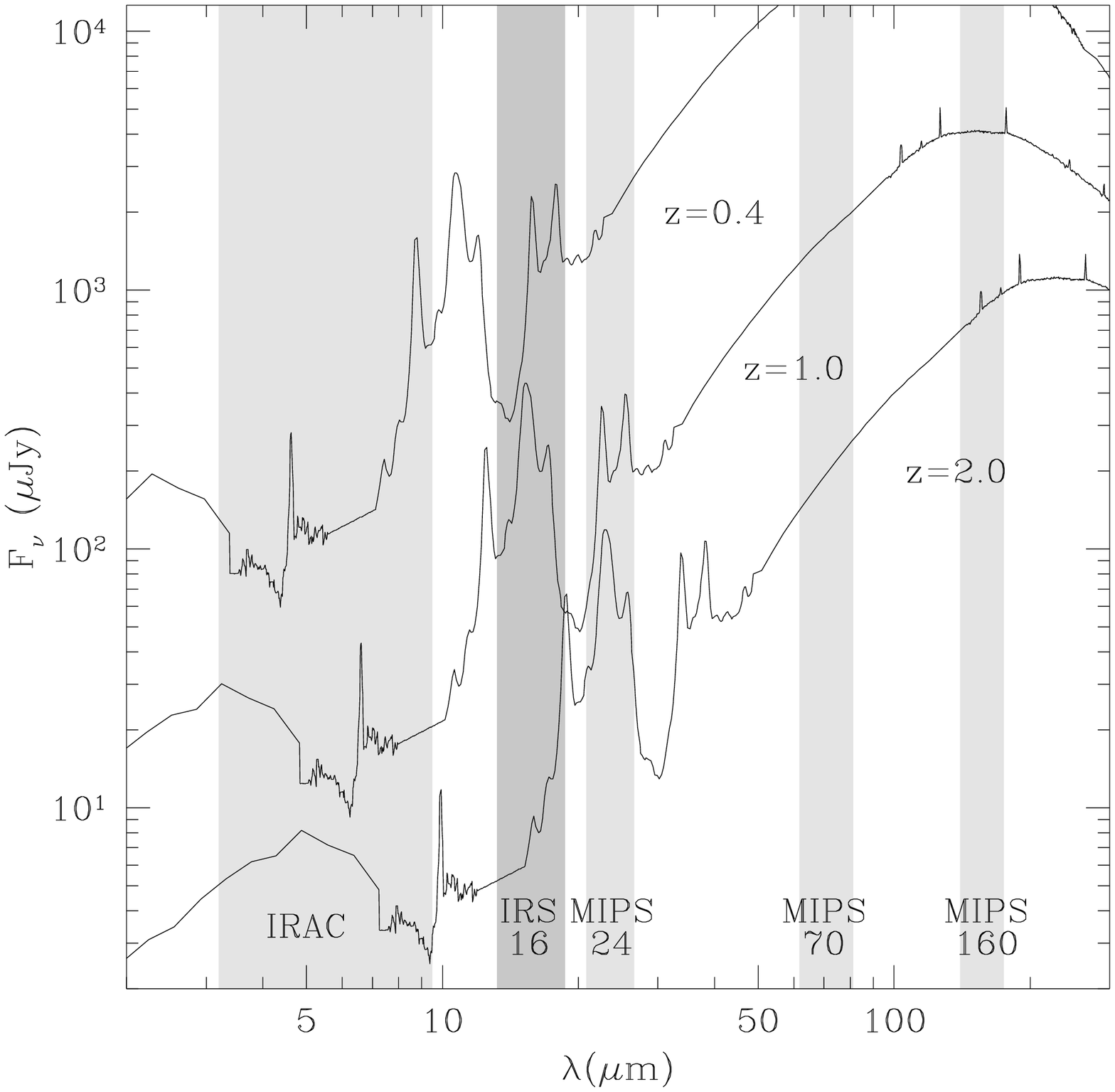}
\includegraphics[angle=0,width=8cm,clip=true,angle=0]{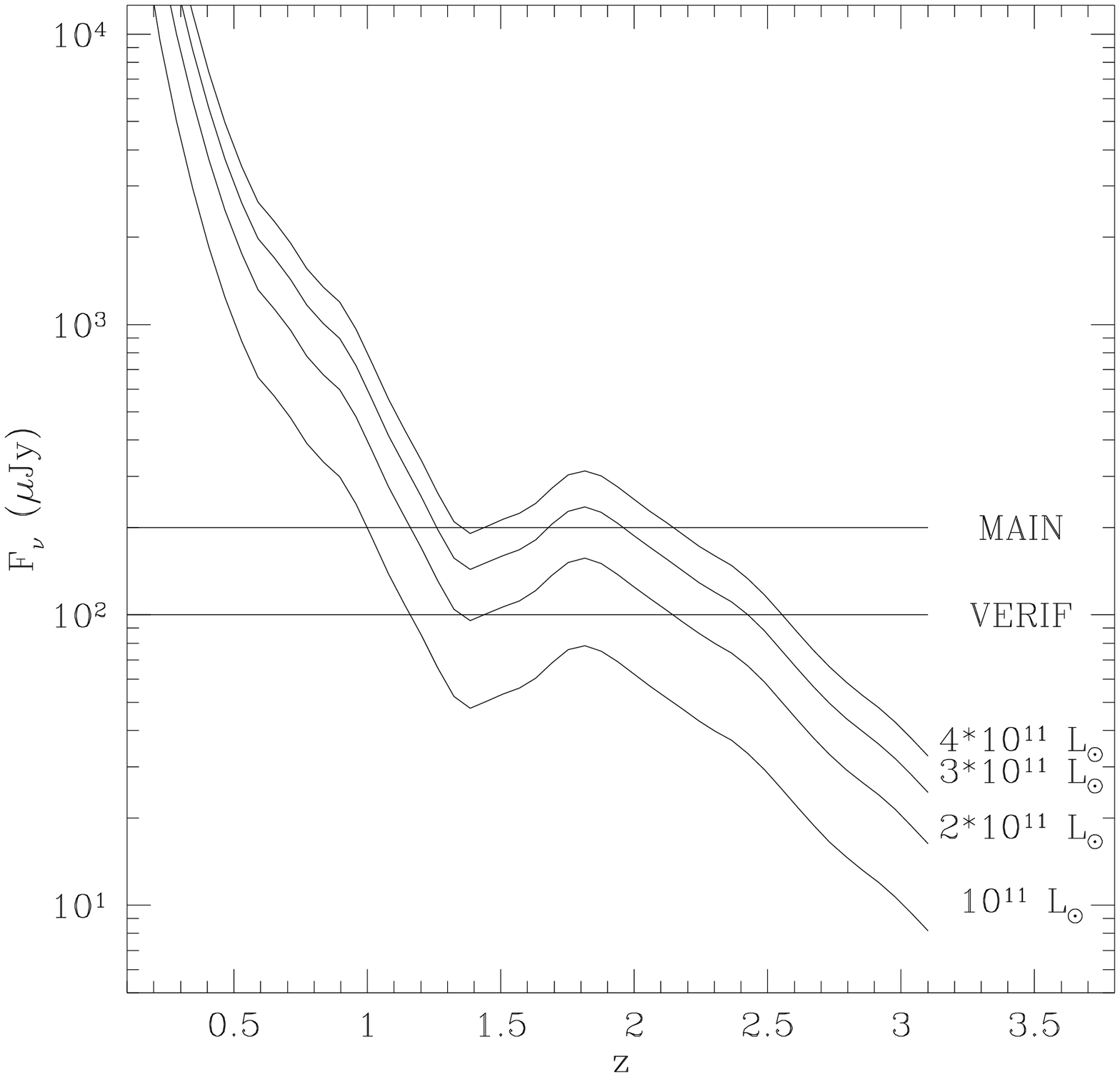}
}
\caption{{\em Left:} A typical LIRG ($L_{IR} = 10^{11} L_{\odot}$) as
seen at three different redshifts by different Spitzer detectors (SED
of M82 renormalized). {\em Right:} 24$\mu$m flux of the same galaxy
with different luminosities as a function of the redshift. Horizontal
bars show the sensitivity limits of the main and verification
survey. In the verification region, LIRGs can be detected up to a
redshift of 2.5.
\label{fig:filter}
}
\end{figure*}

The choice of FLS field was dictated by observational constraints. It
corresponds to the region in the continuous viewing zone of the
satellite with the lowest cirrus contamination (1.1 MJy/sr at 100
$\mu$m, Schlegel et al.\ 1998). In this way, the FLS was observable in
the first scientific campaign independent of the launch date.  Since
the field is a new cosmological field, a series of ancillary
observations were obtained to support the Spitzer data. 
Deep optical data in the R band and VLA data at 20 cm are already
publicly available (Fadda et al., 2004 and Condon et al., 2003).  The
field is part of the Sloan Digital Sky Survey (SDSS; Hogg et al., in
prep.)~\footnote{http://ssc.spitzer.caltech.edu/fls/extragal/sdss.html}.
Moreover, the field has been observed in i' and g' with the 200in
Palomar LFC camera (Glassman et al., in prep.), u' and g' with the CFH
Megacam (Shim et al., 2006), in J and Ks with KPNO/FLAMINGOS (Choi
et al., in prep.) and redshifts have been collected with WIYN/Hydra
(Marleau et al., 2003) and Keck/DEIMOS (Choi et al., 2005).
Finally, to better study the impact of the cirrus emission on
extra-galactic surveys, a HI image of the field has been obtained
with the Green Bank Telescope (Lockman \& Condon, 2005).

In this paper, we discuss in detail the reduction of the data as
obtained from the Spitzer archive, providing a catalog 
and identifications of the sources  with optical and radio
counterparts.  Section~2 presents the observations and the set of data
obtained.  Section~3 describes the steps in the data reduction. The
source extraction is discussed in Section~4 which also includes
a study of the photometric calibration based on stars detected in
the FLS field (calibration factor and PSF estimation). The cirrus
foreground is presented in  Section~5 and the main clouds are
identified by comparison with the HI observations.  Section~6
describes the identification of optical and radio counterparts.
Finally, in Section~7 images and catalogs are presented and a
summary is drawn in the final section.

\section{Observations}



\begin{deluxetable}{ccccr}
\tabletypesize{\scriptsize}
\tablecaption{Description of the observations.\label{tbl:aors}}
\tablewidth{0pt}
\tablehead{
\colhead{AOR key} & \colhead{Date of Start}   & \colhead{Date of End}   &
\colhead{DCEs $\times$ legs} & \colhead{Tot Exp Time}
}
\startdata
3865856 & 2003-12-09 16:56:13 &2003-12-09 18:13:46 &101 $\times$ 10 & 3696.2 s \\
3866112	& 2003-12-09 18:22:45 &2003-12-09 19:40:15 &101 $\times$ 10 & 3696.2 s \\
3866368	& 2003-12-09 20:28:46 &2003-12-09 21:46:23 &101 $\times$ 10 & 3696.2 s \\
3866624	& 2003-12-09 21:55:22 &2003-12-09 23:12:56 &101 $\times$ 10 & 3696.2 s \\
3863808 & 2003-12-09 13:38:51 &2003-12-09 15:48:44 &301 $\times$ 6  & 6621.7 s \\
3864064 & 2003-12-10 06:54:29 &2003-12-10 09:28:46 &301 $\times$ 7  & 7725.3 s \\
3864320 & 2003-12-10 15:44:45 &2003-12-10 18:18:58 &301 $\times$ 7  & 7725.3 s \\
3864576 & 2003-12-10 23:50:42 &2003-12-11 02:00:31 &301 $\times$ 6  & 6621.7 s \\
3864832	& 2003-12-10 04:14:42 &2003-12-10 06:24:35 &301 $\times$ 6  & 6621.7 s \\
3865088	& 2003-12-10 12:39:49 &2003-12-10 15:14:07 &301 $\times$ 7  & 7725.3 s \\
3865344	& 2003-12-10 18:49:38 &2003-12-10 21:23:53 &301 $\times$ 7  & 7725.3 s \\
3865600	& 2003-12-11 02:30:34 &2003-12-11 04:40:22 &301 $\times$ 6  & *6618.0 s \\
 \enddata
\tablenotetext{a}{The asterisk indicates that one DCE was not transmitted.}
\end{deluxetable}


The observations of the FLS field were carried out using 12 AORs
(Astronomical Observation Request) in three days of the first MIPS
scientific campaign (2003, December 9-11) in nominal operations.
The {\em main} field (2.5$\times$2 sq. deg.) was surveyed with 8 AORs to obtain an
average effective integration time of 84 seconds per sky pixel. Four
AORs were used to survey the {\em verification} region
(1.$\times$0.5 sq. deg.) obtaining an effective integration time of
426 seconds per sky pixel.

Compared to other extragalactic MIPS surveys, the main FLS survey is
close to the six SWIRE surveys (Lonsdale et al. 2003) in terms of sky
coverage (5. sq deg compared to 4.8-11 sq deg of the SWIRE surveys)
and sensitivity (1.4 lower than that of the SWIRE surveys).  The
verification survey is similar to the GTO surveys (P.I.: Rieke, in the
CDFS, XMM, HDFN, and Lockman Hole fields) which cover between 0.4 and
0.67 sq. deg. and are 1.4-2 times more sensitive.  The GOODS
ultra-deep surveys (Dickinson 2001) cover an area ten times smaller
with a sensitivity nine times better than that of the FLS verification
survey.

As reported in Table~\ref{tbl:aors}, every AOR is a scan composed of a
variable number of legs. During a scan leg, the motion of the
satellite is compensate by the movement of a mirror (called cryogenic
scan mirror, CSM) which freezes the sky during the integration on the
array. Fifty different positions of the CSM are used during each scan.
Along a scan leg, a cycle of 25 different positions is repeated
several times. Another set of 25 CSM positions is used to compensate
for the satellite motion during the return scan leg.  The AORs which
cover the verification fields used a shorter scan leg than those in
the main field and therefore have fewer exposures (or DCEs -- Data
Collection Events) per scan leg with respect to those in the main
field.

IRAC observations (Lacy et al., 2005) covered approximately a square
of $2 \times 2$ square degrees of the area observed with MIPS at
24$\mu$m (see Figure~\ref{fig:surveys}).  MIPS observations at
70$\mu$m and 160$\mu$m (Frayer et al., 2005) cover approximately the
same region observed at 24$\mu$m since they have been obtained at the
same time. They are displaced by 12.2 and 7.6 arcmin, respectively,
with respect to the 24$\mu$ image on opposite sides along the scan
direction.

In Figure~\ref{fig:surveys}, locations of other public surveys
performed in the area are drawn. A wide-field KPNO survey in the R
band (Fadda et al., 2005) covers almost the entire area surveyed at
24$\mu$m. The entire area has been also surveyed by the Sloan
telescope in the five SDSS bands (Hogg et al., in prep.).  Two radio
surveys (20cm with the VLA, Condon et al. 2003 and with the
Westerbrock, Morganti et al. 2004) were done in preparation
for the infrared survey.  Most radio sources with bright optical
counterparts ($R < 20$) have been
spectroscopically followed-up with WIYN/Hydra (Marleau et
al. 2003). Virtually every radio source with a known redshift has a
counterpart at 24$\mu$m.  Finally, a field of 0.12 sq. deg. has been
observed with the Hubble Space Telescope (ACS I-band parallel survey during
NICMOS observations of EROs, Storrie-Lombardi et al., in prep.) in the
region of deep 24$\mu$m data (verification survey).

The passband of the MIPS24 filter is compared to the other Spitzer filters in
Figure~\ref{fig:filter} . It covers the wavelength range 20.7~-~25.8$\mu$m 
(40\% transmission limits) and its effective wavelength is
23.675 $\mu$m.  Luminous infrared galaxies ($L_{IR} > 10^{11}
L_{\odot}$) are common in high redshift surveys (see e.g. Elbaz et
al. 2002) and such galaxies can be seen in our survey up to a redshift
of $\sim$2.5.

\section{Data Reduction}
\label{sct:datared}

We started our data reduction with data obtained from the Spitzer
science archive: basic calibrated data (BCD) and flats computed for
the MIPS campaign.  For this paper we used the products from the SSC
pipeline version S10.5 (May 2004).  The BCDs are individual frames
already corrected for dark, flat and geometric distortions.  The darks
used in S10.5 were computed from a set of observations obtained early
in the mission and the flats were obtained from scientific data of the
entire MIPS campaign. For an optimized reduction we choose to improve
the dark and flat corrections using the same set of data.  This is
possible since our observations are not dominated by background
fluctuations. Therefore, once the sources are masked, the background
is essentially flat.

\subsection{Dark Residuals and Jail-bars}

The 24$\mu$m array has four independent output amplifiers servicing
interleaved columns of arrays. This produces a characteristic
dark-current pattern of four columns repeating across the array.

\begin{figure}[t!]
\includegraphics[angle=0,width=8cm,clip=true,angle=0]{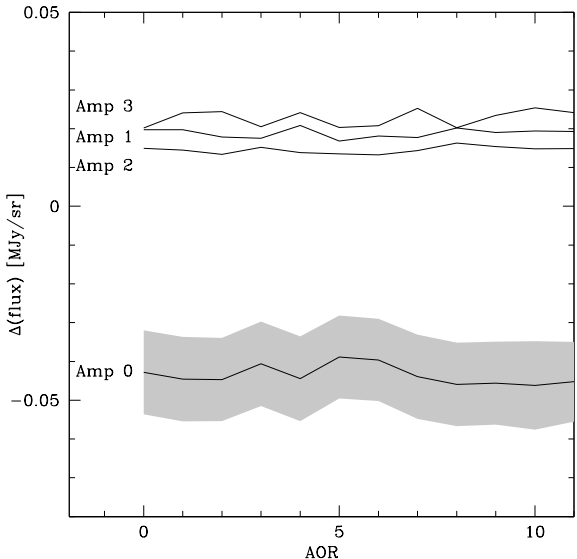}
\caption{Median flux shift of pixels read by the four different
amplifiers in the 24$\mu$m array with respect to the smoothed image,
as a function of the AOR. The median value for each AOR is shown. In
the case of the first amplifier the error bar is also shown.  Data
have been corrected for this effect due to an imperfect dark
subtraction.
\label{fig:dark}
}
\end{figure}

Residuals of a bad dark correction are visible in the archival flats.
To correct for this effect, we de-flattened the BCDs with the archival
flats. These are a set of 50 images, each one corresponding
to a different position of the cryogenic scan mirror (CSM). 

We median filtered every BCD with an 8$\times$8 pixels box to obtain
a smoothed background image and then subtracted from the original BCD.  Then,
masking every pixel affected by cosmic rays, we computed the median
value of flux for the pixels corresponding to each one of the four
amplifiers. The value obtained is the offset required to correct
the dark applied to the BCD. In a second iteration, when the first
final map and a catalog of sources is obtained, sources are also
masked.

As shown in Figure~\ref{fig:dark}, the amplifier connected to the
first column has a flux systematically lower than the other three.
The difference is constant for every AOR, while very small variations
occur among the different AORs. The amplitude of the difference
(approximately 0.07 MJy/sr) corresponds to 1.5 times the noise in the
final image and it is therefore a non negligeable effect.

We computed the median value of the flux for every amplifier during
each AOR and corrected the column flux of each BCD.  After this
additive correction, there are still a few BCDs which have dark
latencies especially after sudden variations in the incident flux, as
after a reset of the detector or after a cosmic ray hit. In Spitzer
jargon these are called ``jail bars'' because typically one out of
every four columns is enhanced with respect to the others. Thus, a
regular pattern of bars appear in the BCD recalling the image of a
prison window.  An additive correction was applied to the BCDs for
which this effect is 5 times greater than the dispersion of the flux
offsets.

In rare cases a few lines are enhanced in the array, probably
because of a glitch in the electronics during the reading of the CCD.
This effect is clearly visible in the HDF-North observations (see
Fadda et al., in prep.). In the case of the present observations no
BCDs were affected. Nevertheless, a few BCDs have a set of lines masked
because of data transmission problems.

\begin{figure}[t!]
\includegraphics[angle=0,width=8cm,clip=true,angle=0]{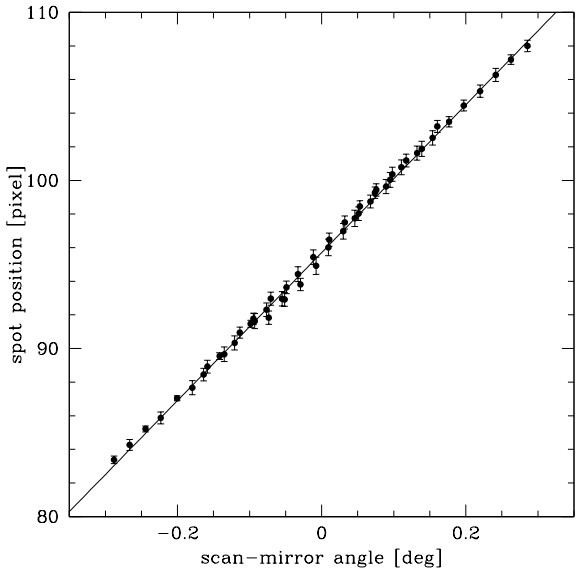}
\caption{Scan-mirror angle versus y-pixel coordinate along the column
116 of the biggest spot due to the pick-off mirror for the 50 different
positions of the scan mirror used during the observations. 
 The overplotted line is the fitted relationship: $y[pix]=95.7+sm[deg]*44.0$.
The model used for computing the scan-mirror position reproduces very
well the actual position of the scan-mirror.
\label{fig:spotpos}
}
\end{figure}

\subsection{CSM Effects on Flat and Illumination}

As previously discussed, the cryogenic scan mirror (CSM) is used to freeze
the image of the sky during the integration since the observatory is 
continuously moving during the observations.

Unfortunately, 
the pick-off mirror has some dirty spots which are projected on the array
in different positions corresponding to the various CSM orientations.
This means that flats have to be computed for each set of frames with
the same CSM orientation.  Moreover, since CSM positions slightly vary
from AOR to AOR and since latencies of bright objects from previous
observations can be visible along an entire AOR, it is advisable to
compute a different set of flats for each AOR. Since the FLS
observations have very long AORs, there are enough frames to compute
flats per each CSM position and per each AOR (36 and 20 in the main
and verification surveys, respectively).

The position of the scan mirror at the beginning of each exposure
found in the header of the BCDs is computed through a model. Since at
least one spot is clearly visible in each frame, we have computed its
position and compared it to the value in the header of the BCD to
check the accuracy of the model predictions. In particular, we have
considered the spot always present along column 116 of the array for
the 50 positions of the CSM used during one AOR. To compute its
position along the column we masked all the pixels affected by cosmic
ray hits and/or sources and fitted a parabola.  The minimum of the
parabola fit gives the position of the spot.  As visible in
Figure~\ref{fig:spotpos}, the position of the CSM computed with the
model reproduces very well the actual position of the CSM.

\begin{figure}[t!]
\includegraphics[angle=0,width=8cm,clip=true,angle=0]{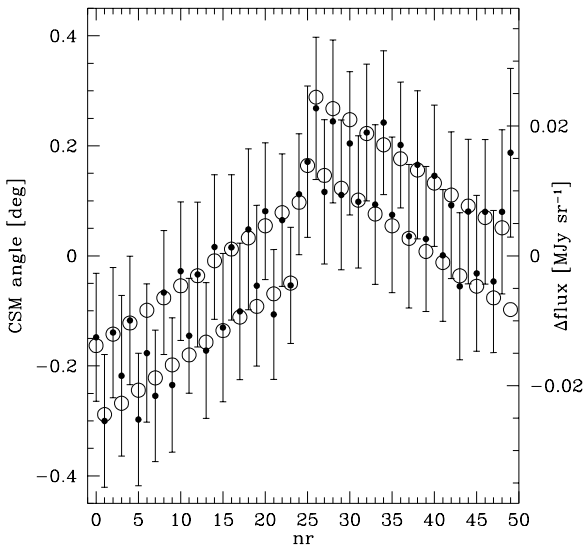}
\caption{Average positions of the scan mirror during an integration
(empty circles) compared to the difference between the frame median
flux and the median flux along the entire scan (full circles with
errorbands) for the 50 different positions of the scan mirror used
during the observations.  The first 25 positions are used during a
forth scan-leg, the second 25 during a back scan-leg.  The flux
differences have been matched to the scan-mirror angles by multiplying
them by 11.8.  The obtained match is very good, except for the last
position where the illumination is uncorrelated to the position of the
scan mirror.
\label{fig:illumination}
}
\end{figure}

A variation of the array illumination which correlates with the CSM
orientation is also clearly visible in the history of the median
fluxes (see Figure~\ref{fig:corrections}).  To better study this
effect, we have considered the median fluxes of frames with the same
CSM position and matched them to the CSM angles
(Figure~\ref{fig:illumination}).  In detail, for all the frames with
the same CSM position we have computed the median flux on the frame,
subtracted the median flux of the entire observation and computed the
average and dispersion of the obtained values.  The best match of the
deviation from the median flux with the CSM angle is obtained by
scaling the CSM angle with the factor 11.8.  The correlation of the
CSM positions with the illumination variation is striking except for
the last CSM position.
We applied a multiplicative correction to the data, since we noticed
in the observations of standard calibrators that the flux of the stars
in the different frames variates with the CSM angle in the same way
as the background flux.

To correct for this effect we therefore used the CSM position
(rescaled with the factor 11.8) for all the orientation but the last
angle for which we accepted the value from the actual measurement.  It
is not clear why the correlation fails for this CSM orientation.
Since the CSM position is correctly modeled, as we can see from
measuring the spot position on the array (see
Figure~\ref{fig:spotpos}), the effect has to be related with another
phenomenon.  Moreover, in other medium scan surveys recently reduced by
the first author, the correlation works perfectly for all the
positions of the CSM.

\subsection{Zodiacal light and long-term transient}

\begin{figure}[t!]
\setlength{\fboxsep}{3.5cm}
\framebox[8cm]{See figure 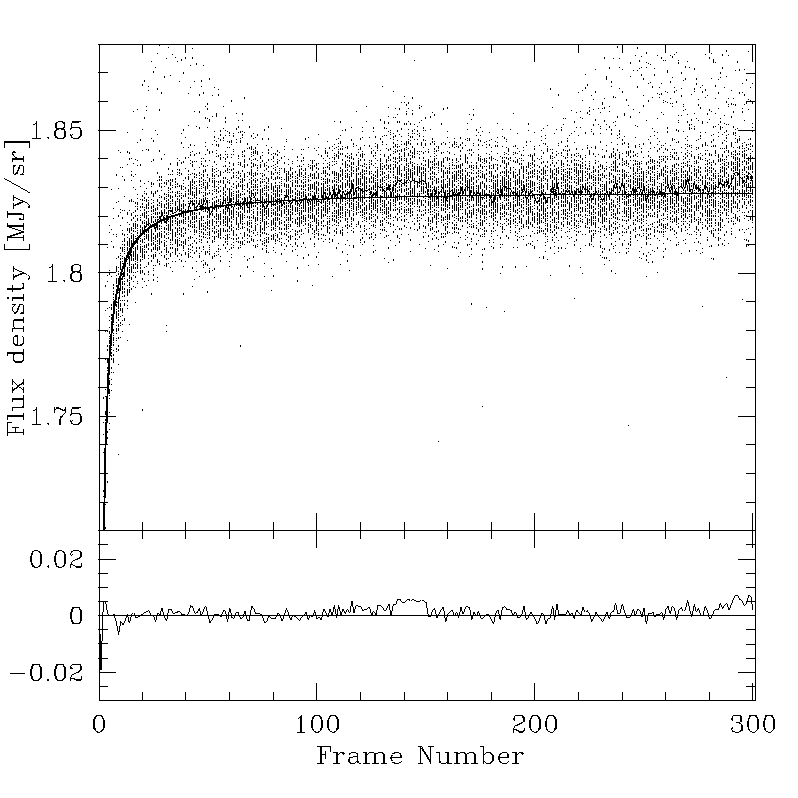}
\caption{Long term drift in the median flux as a function of the
exposure number. A median transient (thin line) is computed by
considering all the scan legs of the main survey (dots). An exponential
function fitting the median behavior is used for the correction (thick line).
The low part of the figure displays the residuals after applying the 
correction to the median line.
\label{fig:ltt}
}
\end{figure}

\begin{figure*}[t!]
\includegraphics[angle=0,width=12cm,clip=true,angle=-90]{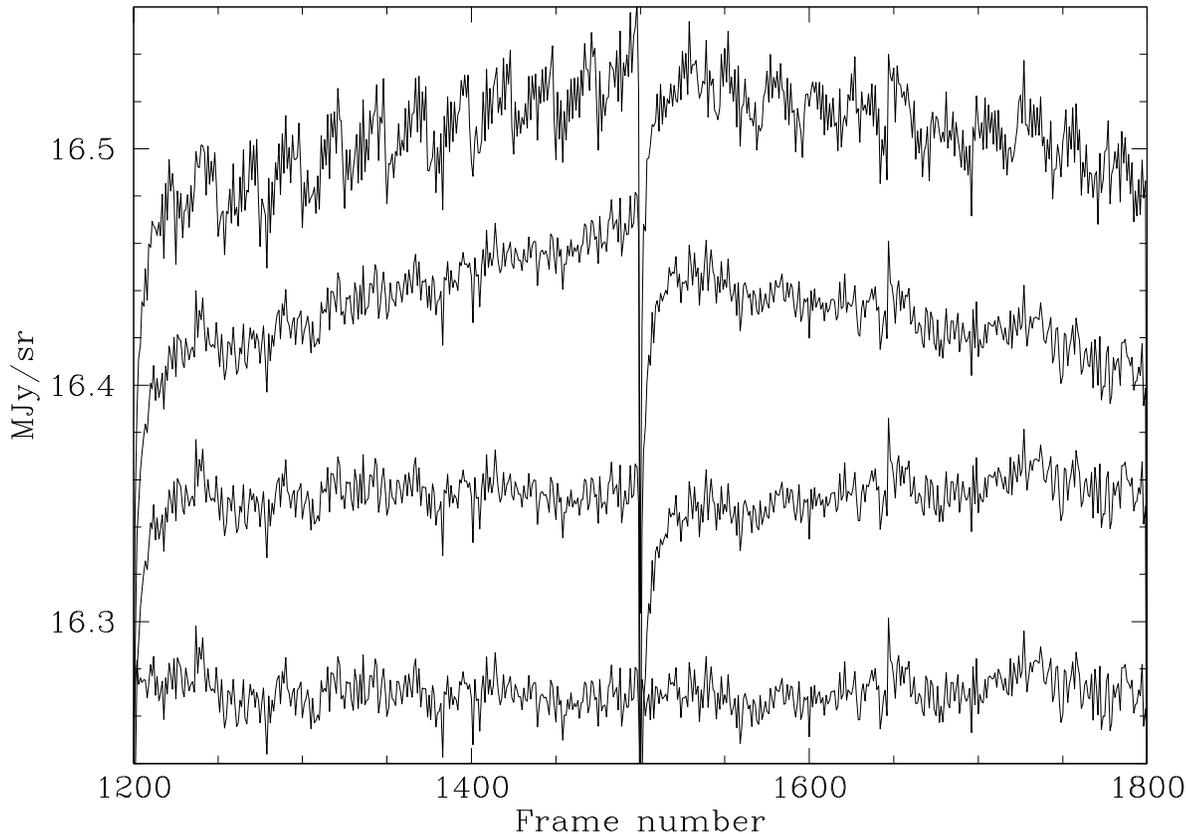}
\caption{The effect of corrections on the median value of the frames
of two main-survey subsequent scan legs. From the
top to the bottom: the original median values of the frames; 
after scan-mirror illumination correction;
after subtracting the zodiacal light;
after subtracting the long-term transient.  
The flux density refers to the uncorrected
signal. The others have been arbitrarily shifted along the y-axis 
to show the effect of the corrections.
\label{fig:corrections}
}
\end{figure*}

After correcting BCD median levels for the CSM-dependent illumination,
there are still three other components that affect their median
fluxes: zodiacal light, a long-term transient and Galactic cloud
diffuse emission.  We have subtracted the variation due to the
zodiacal light by using the Kelsall et al. (1998) model available
through the {\it Spot} software used for planning Spitzer
observations. In particular, we have computed the predicted value of
the zodiacal light for every BCD taking into account its sky position
and the time of the observation.  In this way we are able to model
accurately the variations in the flux along the AORs, as shown in
Figure~\ref{fig:corrections} although the absolute flux obtained is
probably incorrect.

Since we are interested not only in studying point sources in the
field but also in retrieving the variations of the foreground due to
Galactic cloud emission, we tried to model the long-term variation of
the detector. This feature has been already detected with ISOCAM and
in that case an attempt at a correction was made using the redundancy
of the observations ( Miville-Desch\^enes et al., 2000).

In the FLS survey we have the advantage of having several AORs
performed in exactly the same way in a region of sky with low Galactic
infrared foreground.  We are therefore able to study the median
behavior of the long-term transient by considering all the scan legs
of each AOR, since at the beginning of each scan leg the detector is
reset.  Since the scan legs are each done in a different region of sky and
the legs cover over 2  degrees , the median behavior will be
independent of the small-scale variations due to Galactic cloud
emission (the typical size of a cloud is less than half degree).
In Figure~\ref{fig:ltt} all the BCD median fluxes are shown together
after shifting each scan-leg in flux to better
match, in the sense of the minimum $\chi^2$, the scan-leg
with less variations (i.e. less foreground Galactic emission).
An exponential function has been fitted to the points, then the 
matching of the different scan legs have been repeated by shifting
in flux the scan legs to match the fitted curve and refitting the function.
The median behavior is well fitted by the exponential function:
\begin{equation}
F(t)=28.2 e^{-0.04/t(s)}-26.37
\end{equation}
with time in seconds (time elapsed from the first exposure) and
flux units in MJy/sr.
The flux level computed with this function is then subtracted
from each BCD to obtain a final mosaic with a median zero level conserving
at the same time the structure of the Galactic foreground.

 In the case of the verification survey, where the scan legs
are shorter, we have shifted all the scan-leg BCD fluxes to match
the function fitted with the main survey data.
The shifts added to the function values have been subtracted from
the BCDs.

The effects of CSM illumination correction, subtraction of zodiacal
light and long-term transient are shown in
Figure~\ref{fig:corrections} on the median BCD fluxes of a piece of an
AOR. All these corrections are needed to match the backgrounds
measured by different BCDs in the same region of the sky. If the BCDs
are coadded without applying these corrections, the variations in the
background increases the noise in the final image.  As we will see in
section~\ref{sct:cirri}, we can in this way also recover the structure
of the Galactic clouds emitting at 24$\mu$m.

\subsection{Relative Offsets}
\label{sec:rel_off}

Although the pointing of Spitzer is very accurate, errors of the
order of an arcsec in the absolute pointing are expected.
On  top of that, relative offsets are expected during an observation
because the telescope is adjusting its pointing every time it moves.
The biggest errors are therefore expected along the scan direction.
If not corrected, these relative offsets will affect the PSF in the final mosaicked
image.

\begin{figure}[t!]
\includegraphics[angle=0,width=8cm,clip=true,angle=0]{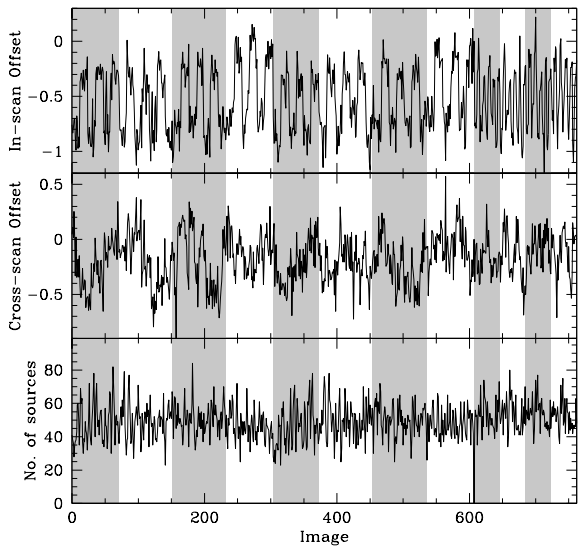}
\caption{Offsets of groups of 25 images with the same telescope pointing
along the in-scan and cross-scan directions in arcseconds with respect
to the r' image of the field obtained by the SDSS. The bottom panel reports the
number of sources with optical counterpart in each image used for computing
the offsets. The grey and white background separates the different AORs.
\label{fig:offsets}
}
\end{figure}

The ideal way to proceed would be to compute the offsets with respect
to an image with accurate astrometry for each BCD. Unfortunately, the
number of sources visible in each BCD is usually not high enough to
compute an offset with a reasonable error. We therefore decided to coadd
every 25 consecutive BCDs which have been observed by moving only the
CSM. We expect in fact that the biggest errors in pointing occur when
the telescope itself has moved.

As a reference image we used the SDSS r' image of the field.  This
observation has the advantage of having accurate astrometry (error $<
0.1$ arcsec) covering the entire field and to be sufficiently deep to
find a high number of optical counterparts of the infrared sources in
every small mosaic.  Typically, we find 50 sources with optical
counterparts in each mosaic.  The offsets are computed along and
perpendicular to the scan direction.  Figure~\ref{fig:offsets} reports
the number of sources used to compute the offsets and the offsets
along the in- and cross-scan directions for every set of 25 BCDs.  The
shadows on the figure separates the different AORs. Considering the
offsets in the in-scan direction, the offsets clearly depend on the
direction of the scan leg. To achieve a better accuracy, one could in
principle apply only two corrections per AOR along the in-scan
direction corresponding to the two directions of the scan.  I chose
not to follow such a procedure because, in some cases, groups of
images show a big deviation from this median behavior. Since, as shown
in the bottom panel of Figure~\ref{fig:offsets}, the offsets are
measured using around 50 sources, it is difficult to believe that
these deviant offsets are simply due to insufficient statistics.  The
overall offsets along the two directions are always less than one
arcsecond.

The astrometry of each BCD has been updated according to the offsets
measured in the two directions.

\subsection{Mosaicking}

A total of 19691 BCDs were obtained in the FLS main and verification surveys.
92 of them have a shorter integration time and were obtained 
at the beginning of each scan leg. Although these frames can
in principle be used, the detector is far from stabilization
during these exposures and typically they are affected by
strong jail-bars. We do not include these frames in our final
image mosaics.

To coadd in a single map the remaining 19599 BCDs we used
MOPEX, a multi-purpose tool developed at the SSC by D. Makovoz (see
Makovoz \& Marleau, 2005) to 
find bad pixels, coadd and extract sources. 
\begin{figure*}[t!]
\includegraphics[width=7cm,clip=true,angle=-90]{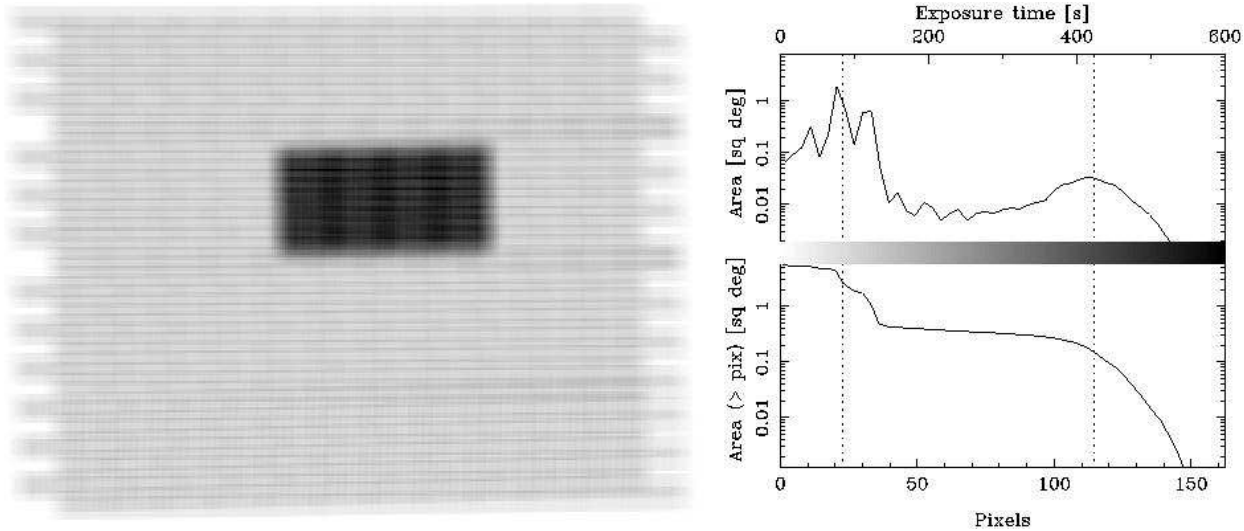}
\caption{{\em Left: } coverage map of the MIPS24 observations. {\em
Right:} the differential (top) and cumulative (bottom) plots of the
coverage versus area. The two vertical dotted lines mark the median
coverage in the main and verification surveys.
\label{fig:coverage}
}
\end{figure*}
The final map has a pixel size of 1.275
arcseconds, half of the original pixel size. This allows us
to exploit the super-resolution due the different
observations taken with shifts of fractional pixels.

Although the SSC pipeline provides a mask of bad pixels
which are damaged or saturated, several other pixels have to
be masked because of cosmic rays. Among the several possibilities
offered by MOPEX to detect these pixels, we choose the most
conservative one which relies on redundancy.

As visible in Figure~\ref{fig:coverage}, the typical coverage is 22.6
and 114.4 in the main and verification surveys, respectively.  This
gives enough values to reject deviant pixels affected by cosmic rays
with MOPEX. The rejection has been done in a conservative way using a
top threshold of 5$\sigma$ and a bottom threshold of 3$\sigma$.  Using
a too low top threshold is dangerous, because this can mask the
brightest pixels of bright objects which usually have larger
variance, or the faint wings of the PSF which are at the noise level.

Once the new masks are produced, a mosaic and a list of sources can be
obtained. In order to improve our reduction, we used the improved
masks and the list of sources to redo darks, flats and further
corrections based on the median flux of the BCDs by masking sources
and bad pixels.

In case of bright sources, latencies are visible in the BCDs taken
after the observation of the bright source. In the case of our field,
only one source is bright enough to produce latencies and therefore
ghost sources in the final mosaic if these are not corrected.
We corrected this case empirically as explained in section~\ref{sec:psf}.
We then reprojected the corrected BCDs to obtain a final mosaic.
The namelist used for the projection is available through the SSC science
archive (see Section~\ref{sct:dataprod}).

\section{Data Analysis}

\subsection{Source Extraction}

The point source extraction was performed using the MOPEX software
(Makovoz \& Marleau 2005).  The extraction was done in single frame
mode, that is using the same image for detection and fitting of point
sources.  The procedure used consists of the following steps.

In the first step, a sample of isolated and bright point sources was
selected to estimate the empirical point response function (PRF) from 
the mosaic. The module {\sl
apex\_1frame.pl} was run using a theoretical PRF produced by the
STinyTim, the {\it Spitzer} version of the Tiny Tim Point Spread
Function (PSF) modeling program\footnote{Developed for the {\it
Spitzer} Science Center by John Krist}.  The mosaic is filtered
with the simulated PRF to produce a point source probability (PSP)
image which is then used for detection. This filtering process
enhances the point sources that match the input PRF while smoothing
out noise features.  The fitting was done without active deblending and
only the brightest isolated sources (flux $> 5 \times 10^3$ $\mu$Jy)
with the lowest $\chi^2$ ($\chi^2/dof < 30$) are kept. A total of 27
sources were selected for the PRF estimation.
\begin{figure*}[t!]
\includegraphics[angle=0,width=16.8cm,clip=true,angle=0]{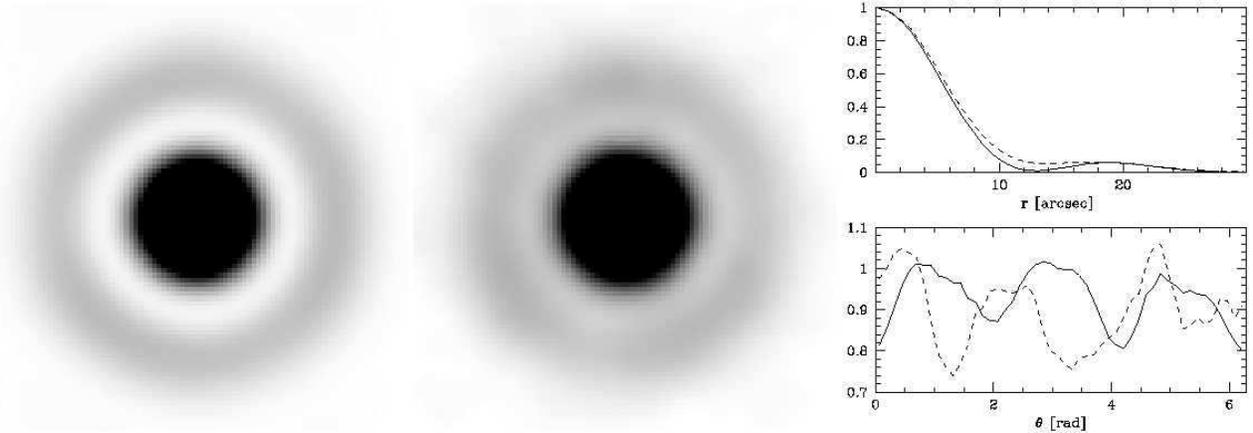}
\caption{Theoretical (left) and empirical (center) PRF evaluated by
stacking 27 bright isolated point sources. On the right, averaged
radial profile (top) and averaged angular profile of the first
Airy ring (bottom) for the two PRF. The radial profiles are normalized
with respect to their maxima, while the angular profiles are normalized
with respect to their median values. Solid and dashed lines refer to
the theoretical and empirical PRFs, respectively. 
The empirical PRF looks blurred with respect to the theoretical one
and has a different angular profile.}
\label{fig:prf}
\end{figure*}

The PRF was estimated in the second step using the module {\sl
prf\_estimate.pl}. In our case, we selected a postage stamp size of
$35\times 35$ pixels and the radius of 11 pixels which contains the
first Airy ring.  The PRF flux is normalized within that radius and
therefore an aperture correction needs to be applied to the estimated
fluxes (see section~\ref{sec:psf}).  Theoretical and empirical PRFs
are shown in Figure~\ref{fig:prf}.  The PRF estimated from the data
appears blurred with respect to the theoretical PRF. The different
angular profile is due to differences between the model and the actual
telescope, while the blurring is due to pointing uncertainties and
sampling. The observed radial profile can be recovered by
taking into account how the image is constructed, as done in the study
of aperture corrections (see Section~\ref{sec:psf}). The angular
structure depends on the details of the optics.

The first bright Airy ring around bright point sources causes many
false detections in the mosaic image at faint flux levels. Therefore, our
third step was to create a mosaic image where all the point sources
with SNR $>$ 20 in the point source probability image have had their
Airy rings removed. Using {\sl apex\_1frame.pl} with the empirical
PRF, a second extraction at SNR $>$ 20 was made obtaining a total
of 1266 sources. A residual image was created using the module 
{\sl apex\_qa.pl} using a modified PRF with the region inside the
first minimum set to zero.

\begin{figure}[t!]
\setlength{\fboxsep}{3.5cm}
\framebox[8cm]{See figure 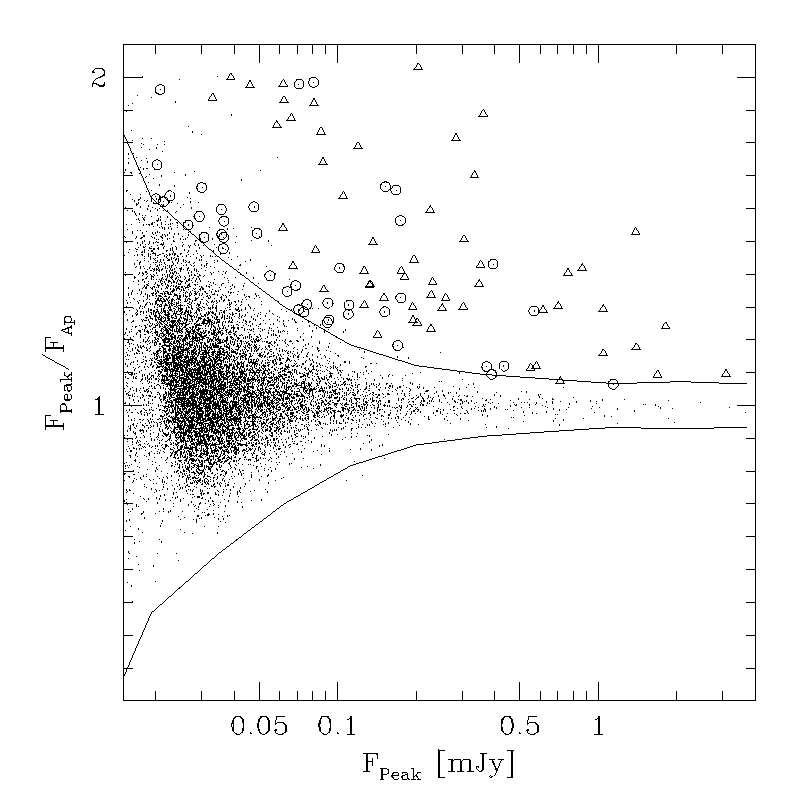}
\caption{Ratio between peak and 5'' aperture fluxes. Points higher than
the 3-$\sigma$ errorband are candidate extended sources. The triangles are
the extended sources found with {\it run\_bright\_detect} module. 
The points encircled appear extended in the optical images.
The two negative outliers correspond to two sources on the border for
which the aperture misses part of the flux.}
\label{fig:fluxratio}
\end{figure}

\begin{figure}[t!]
\includegraphics[angle=0,width=8.4cm,clip=true,angle=0]{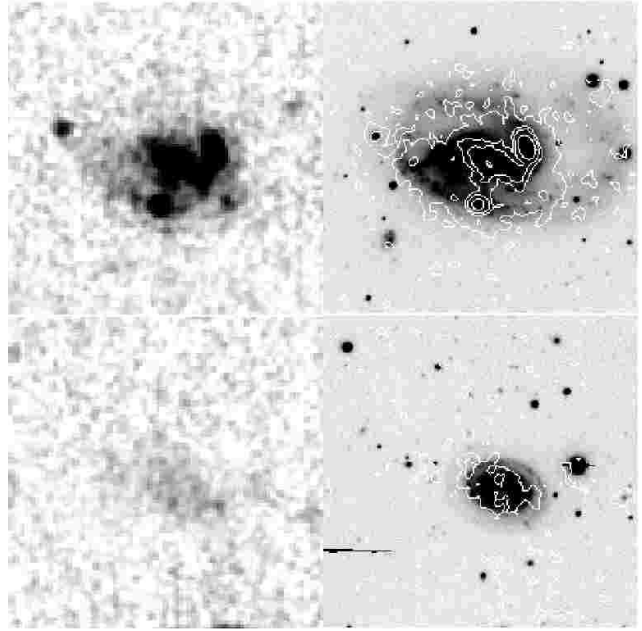}
\caption{Two examples of extended 24$\mu$m sources: on the left the 24$\mu$m image
and on the right the R image with isocontours at the 0.1,0.3,0.5 and 1. MJy/sr levels.
The images have a size of 2$\times$2 sq. arcmin. and show a bright galaxy with 
several bright knots on the top and a low surface-brightness galaxy at 24$\mu$m
on the bottom. The bright galaxy has been identified with the {\sl
run\_bright\_detect} module while the second one with the flux ratio diagram.
}
\label{fig:extexamples}
\end{figure}

\begin{deluxetable}{cccccccc}
\tabletypesize{\scriptsize}
\tablecaption{Extended Sources. (First 10 entries) \label{tbl:extsrc}}
\tablewidth{0pt}
\tablehead{
\colhead{Name} & \colhead{RA}   & \colhead{Dec}   &
\colhead{Flux[24$\mu$m]} & \colhead{Flux[20cm]}& \colhead{R} & \colhead{z} & \colhead{Notes}\\
\colhead{(IAU)} & \multicolumn{2}{c}{(J2000)}   &
\colhead{[mJy]} & \colhead{[mJy]} & \colhead{[mag]} & &
}
\startdata
J170850.7+601144& 17:08:50.744& +60:11:44.11&     7.44&     1.119& 15.13& 0.0667 &  \\
J170855.9+601114& 17:08:55.957& +60:11:14.93&    59.04&     2.230& 15.49&    -   &  \\
J171119.6+602619& 17:11:19.621& +60:26:19.20&     4.29&     0.530& 15.96& 0.0741 &  \\
J171140.1+595944& 17:11:40.194& +59:59:44.79&    31.56&     0.435& 13.61& 0.0167 &  \\
J171226.0+603024& 17:12:26.040& +60:30:24.69&     3.55&     1.283& 15.86& 0.1070 &  \\
J171247.7+592705& 17:12:47.736& +59:27:05.19&     4.18&     0.565& 16.46& 0.1372 &  \\
J171239.4+584147& 17:12:39.482& +58:41:47.18&     9.01&     0.940& 17.50& 0.1653 &  \\
J171336.0+603006& 17:13:36.010& +60:30:06.47&     5.39&     0.231& 20.77&    -   &  \\
J171306.9+591923& 17:13:06.903& +59:19:23.55&     8.82&     0.532& 15.07&    -   &  \\
J171359.5+592017& 17:13:59.529& +59:20:17.03&     4.00&       -  & 15.58& 0.0544 &  \\
\enddata
\tablenotetext{a}{OOI stands for ``out of image''.}
\end{deluxetable}

The detection was then done on the ringless image with a much lower
detection threshold.  A total of 98779 sources were detected.
Finally, the fitting was done with passive and active deblending on the
original mosaic using the list of sources detected on the ringless
image.  A total of 42555 sources ($\sim 1.0 \times 10^4$ sources per
square degree) with SNR $>$ 3 were extracted.
In the final catalog, we considered only sources with SNR $>$ 5 to
have a highly reliable list of sources.

The module {\sl apex\_qa.pl} was used again to obtain an image without
point sources to better estimate the flux of extended sources and to
identify the emission from molecular clouds (see following sections).

\subsection{Extended Sources}

A few extended nearby objects are present in the observed field (see 
Figure~\ref{fig:extexamples}).
These are resolved by MIPS at 24$\mu$m and we can therefore not
extract them in the same way as the point sources.  
\begin{figure*}[t!]
\hbox{
\includegraphics[angle=0,width=8.4cm,clip=true,angle=0]{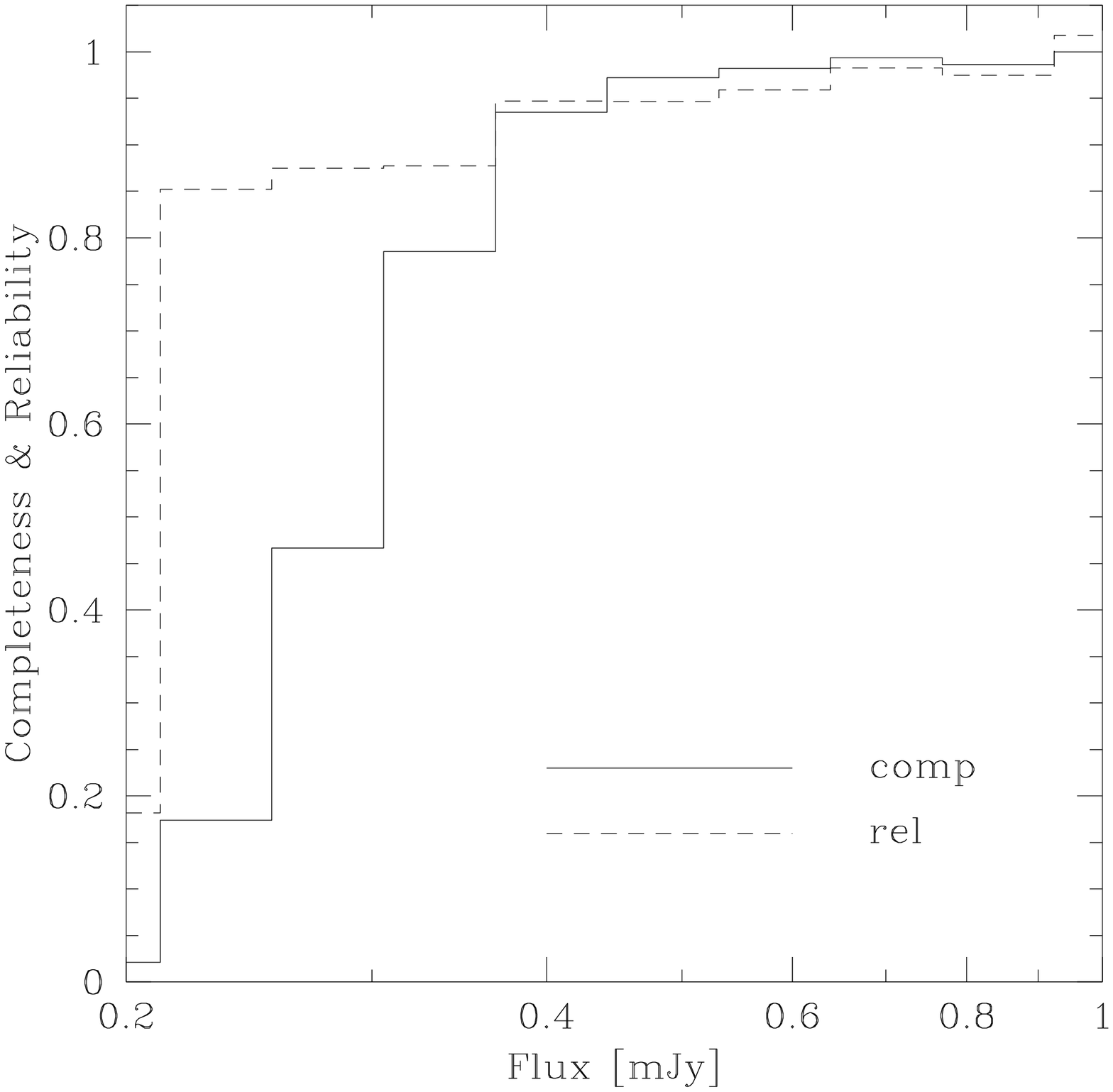}
\includegraphics[angle=0,width=8.4cm,clip=true,angle=0]{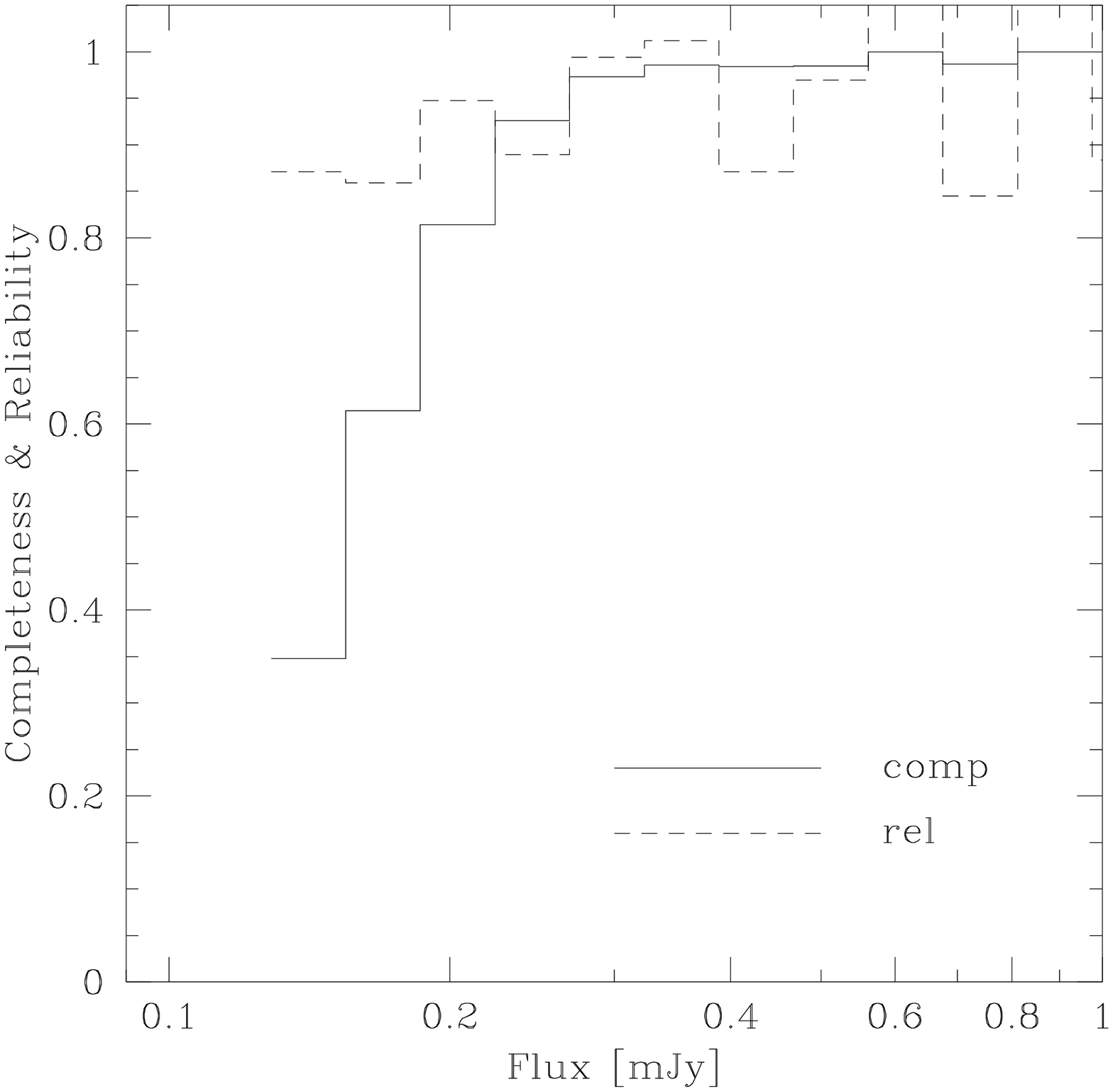}
}
\caption{ Completeness and reliability of the extracted sources at the
5$\sigma$ level in the main (left) and verification (right)
regions. According to our simulations, sources brighter than 0.21 and
0.12 mJy are highly reliable and the surveys are 50\% complete at 0.3 and 0.15
mJy in the main and verification regions, respectively.  }
\label{fig:cr}
\end{figure*}
Most of the extended sources were identified by running the MOPEX
module {\sl run\_bright\_detect} and selecting sources with SNR $>$ 20
and ellipticity $>$ 0.1.  A total of 58 extended sources were
identified in this way.  To find other possible extended sources still
undetected with the previous technique, we compared the flux in the
central pixel with the flux inside a radius of 5 arcsec as in
Figure~\ref{fig:fluxratio}. If the source is extended, the flux is
less concentrated than that of point sources so that the expected
ratio between aperture and peak flux is higher than that of point
sources.  We did not consider in this search all the sources with
companions within a radius of 8 arcsec, since the aperture flux is
affected by the companion source. Usually, sources within this
distance have been deblended by the extraction algorithm used or have
been already identified as extended sources with the {\sl
run\_bright\_detect} module.  We inspected visually the optical
counterpart of each of the extended source candidates with ratio
higher than the 3$\sigma$ errorband.  63 of them belong to extended
objects.  Most of the others are either undeblended objects or low SNR
sources with very uncertain fluxes.  Two negative outliers were found
at the border of the image since the aperture used for computing the
flux misses part of their flux.  All the sources of the point source
catalog falling on extended sources were then removed from the point
source catalog and a new residual image was obtained.  The value of
the total flux was determined by doing aperture photometry on the
residual image with apertures large enough to contain the whole
sources.  The same apertures were masked to obtain a bad pixel list
for the residual image and the masked region were interpolated.  These
regions cover a total area of less than 0.015 sq. deg., that is 0.27\%
of the total area of the survey.  The photometry of the brightest star
in the field has been also evaluated inside an aperture. Although
technically this is a point source, the PRF that we used was not large
enough to do a correct fit and we evaluated the total flux using
aperture photometry.  The first ten entries of the extended sources
list are given in Table~\ref{tbl:extsrc}. The Table reports also the
20cm flux of the VLA counterpart and the R magnitude of the optical
counterpart. Redshifts measured by SDSS are also reported, if
available.

\subsection{Completeness and Reliability}
\begin{figure}[t!]
\setlength{\fboxsep}{3.5cm}
\framebox[8cm]{See figure 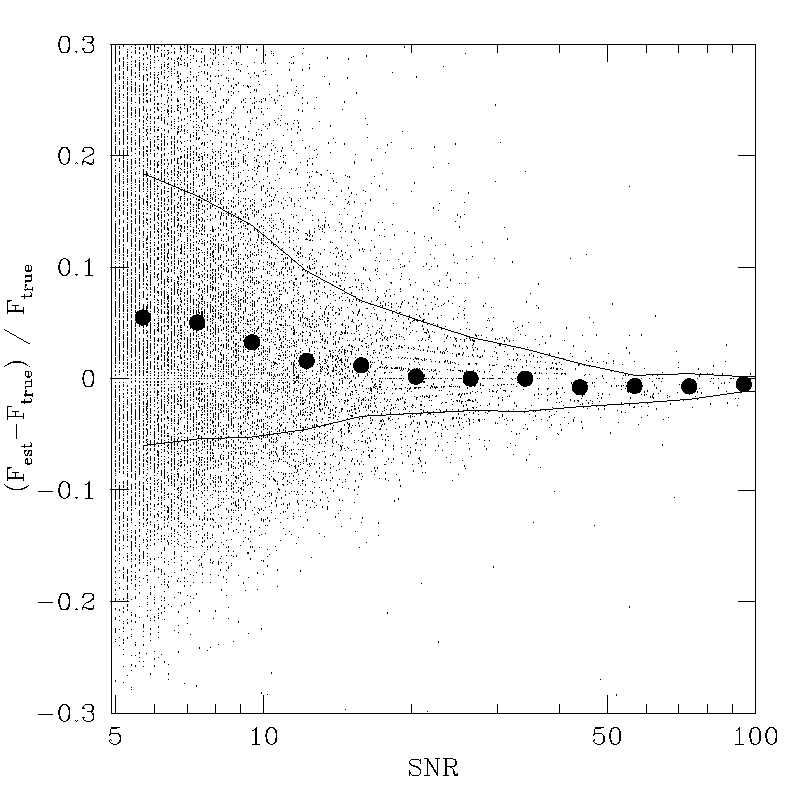}
\caption{Difference between true and estimated fluxes versus SNR for
sources detected in the simulation. Points
correspond to the sources. The big dots correspond to the median difference. 
The two lines mark the 1$\sigma$ error band.
The error increases towards low SNR values. Moreover, at low SNR values
the estimated fluxes are biased towards higher values since sources
on positive fluctuations of the noise are preferentially detected.
\label{fig:fluxerr}}
\end{figure}
\begin{figure}[t!]
\setlength{\fboxsep}{3.5cm}
\framebox[8cm]{See figure 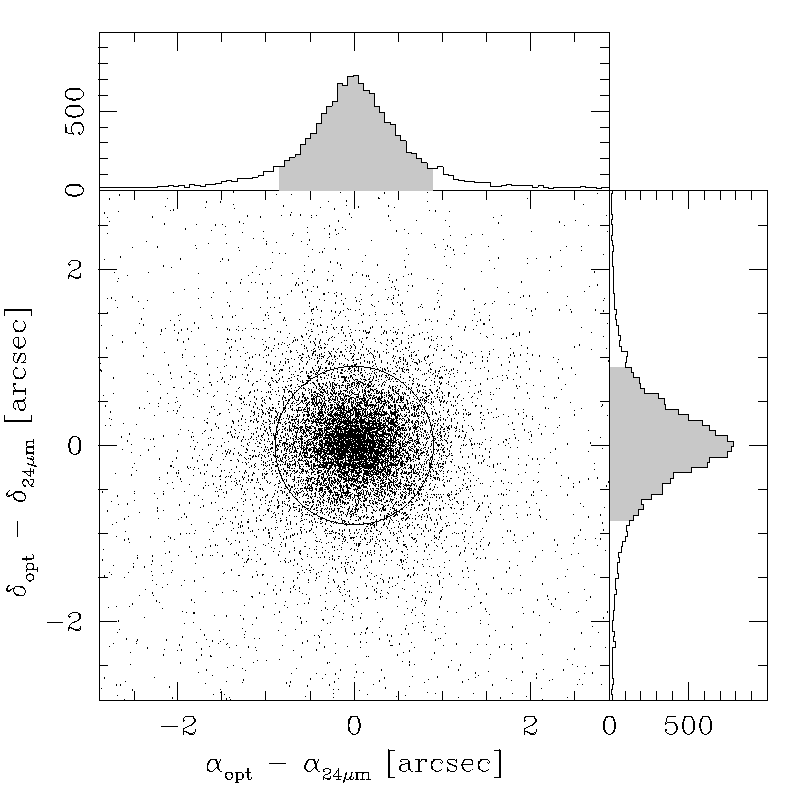}
\caption{Differences in
position between the infrared sources and their optical counterparts. 
68\% of the counterparts are within a 0.9 arcsec
radius circle (see also shaded regions in the histograms).}
\label{fig:poserr_opt}
\end{figure}

To estimate the completeness and reliability of our point source
catalogs we performed simulations reinserting the extracted sources
into the residual image.  In particular, we divided the image in two
regions according to the coverage: (a) coverage greater than 37 images
(roughly corresponding to the verification survey, see
Figure~\ref{fig:coverage}) and (b) coverage between 10 and 37 images
(the main survey except the noisy borders).  Sources extracted (at the
3$\sigma$ level) in one of these region were reinserted in the same
region at random positions using the empirical PRF estimated for the
extraction with the module {\em apex\_qa.pl}.  The regions occupied by
extended sources have been avoided.

In this way, no assumptions about counts and noise structure were
made.  Because of the imperfect subtraction of the sources, the noise
is slightly higher than in the real case.  The extraction of the
simulated sources was done in the same way used in the case of the
real image.  Inserted and extracted sources have been matched
considering a maximum distance of 5 arcseconds and a difference in
flux of 30\%. The constraint in the flux helps in rejecting cases where
a source has been incorrectly deblended.

For the completeness and reliability, we considered only sources
detected with $SNR > 5$ since this is the limit chosen for our
catalog in order to have a high reliability.  As visible in
Figure~\ref{fig:cr}, the reliability is higher than 85\% to the flux
limits of our catalog (0.21 and 0.12 mJy for the main and verification
surveys, respectively).  At the 50\% completeness limits, 0.3 and 0.15
mJy in the main and verification regions, respectively, the reliability
is around 90\%.

Finally, since some sources in our image could be transient sources,
typically asteroids, we obtained two mosaics using images taken at
different times (two consecutive legs of each scan) and subtracted
one from the other to search for residual objects.  No objects were
found at the 5$\sigma$ level.
\begin{figure*}[t!]
\hbox{
\setlength{\fboxsep}{3.5cm}
\framebox[8cm]{See figure 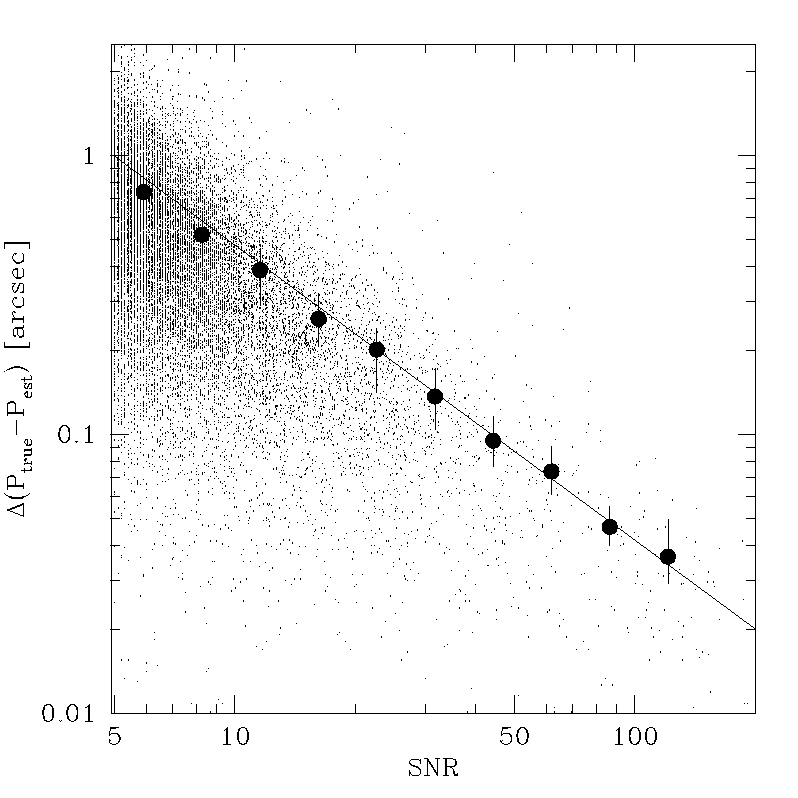}
\framebox[8cm]{See figure 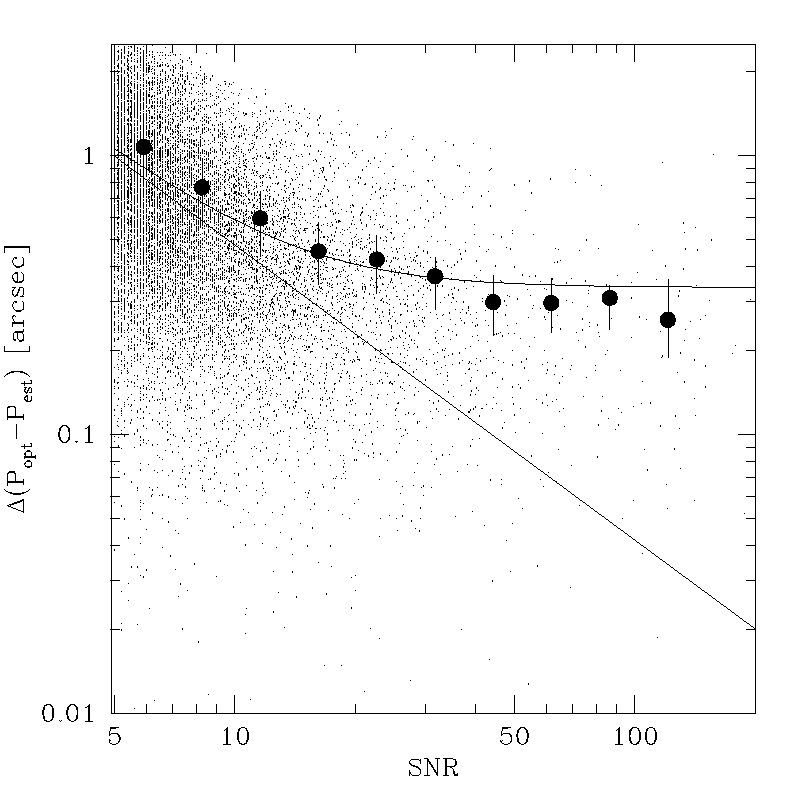}
}
\caption{Left: distance between true and estimated positions versus
SNR for simulated sources. Points correspond to the
sources. The big dots correspond to the 1$\sigma$ error (distance
within which one finds 68\% of the counterparts). The error bars mark the
50th and 80th percentiles of the distribution of points.  The line
shows how the error depends on the SNR value. Right: distance between
infrared sources and their optical counterparts. The 1-$\sigma$ error
can be fitted by summing in quadrature the error computed from simulations (straight
line) with the pointing error (0.3 arcsec) and the optical error (0.1 arcsec).
}
\label{fig:poserr}
\end{figure*}

\subsection{Photometric and Astrometric Accuracy}

Photometric and positional accuracies  are assessed using the same
simulations performed to estimate completeness and reliability.

 The photometric errors come from uncertainties in the extraction and
in the flux calibration.  The uncertainty in the flux calibration is
less than 5\% (Gordon et al., 2006).  To evaluate the uncertainty in
the extraction, we considered the plot with the difference between the
true and measured flux normalized to the true flux versus the SNR of
each detected source (see Figure~\ref{fig:fluxerr}).  The big dot in
every SNR bin corresponds to the median value of the detected points.
Since low SNR sources tend to be detected if they are on top of
positive fluctuations of the noise, this effect creates a bias in the
median estimated flux below $SNR = 10$. In our simulations this effect
is enhanced because of the presence of residuals due to the imperfect
subtraction of the real sources. Since this effect can be corrected
only statistically, we did not apply any correction to the fluxes of
the single sources.  The 1$-\sigma$ error band is also shown on the
plot (region comprised between the 16th and 84th percentiles). The
error increases from 5\% percent at SNR$=20$ up to around 15\% at low
SNR values. The errors reported in the catalogs are computed according
to this curve.

Errors in the astrometry of the 24$\mu$m sources come from the
measurement on the image and from the pointing uncertainty.  We
evaluated the errors made in the extraction process through our
simulations.  As shown in Figure~\ref{fig:poserr}, this error depends
on the SNR of the source. It is typically around 0.8 arcsec for
sources with SNR less than 7 and becomes negligible beyond $SNR =
20$. In the case of bright sources, in fact, more pixels are seeing
the source and this improves the estimate of the center of the PSF.
As explained in section~\ref{sec:rel_off}, we corrected the astrometry
of the BCDs taking as a reference the SDSS image of the field. 
\begin{figure*}[t!]
\includegraphics[angle=0,width=16.8cm,clip=true,angle=0]{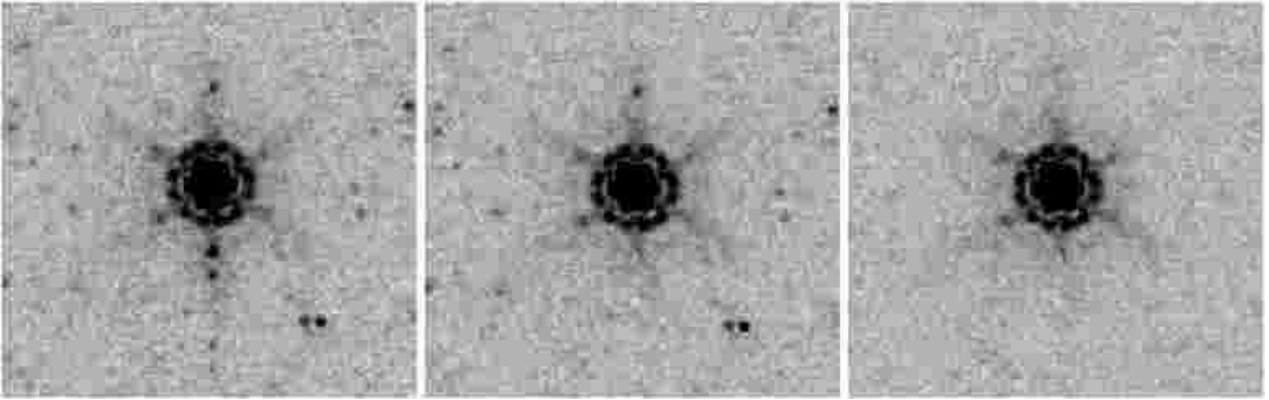}
\caption{From left to right, the image before and after the latency
correction and after the subtraction of surrounding sources. Four
ghost sources below the star are visible in the first image on the
left.  The field shown (4.7'$\times$4.45') roughly corresponds to that
used in the study of the PSF profile.  }
\label{fig:latency}
\end{figure*}
Thanks to this, we do not have any relative offset between the
position of our sources and their optical counterparts (see
Figure~\ref{fig:poserr_opt}).  The pointing error can be computed from
the dispersion in the matching between infrared sources and optical
counterparts during the offset correction.  This is around 0.3
arcseconds. As visible in the right part of Figure~\ref{fig:poserr},
the error computed by matching sources with their optical counterparts
is asymptotically converging to this uncertainty. Considering that the
optical uncertainty is around 0.1 arcseconds, the error in the
matching with the optical sources is dominated by the astrometric
error of the infrared sources.  We can obtain a formula to compute the
astrometric error as a function of the SNR:
\begin{equation}
\sigma^2 = \sigma_{ext}^2+\sigma_{point}^2,
\end{equation}
where the pointing error $\sigma_{point} = 0.3''$ and the error due to
the extraction $\sigma_{ext}$ is estimated from our simulations:
\begin{equation}
\log(\sigma_{ext}) = 0.74 - 1.06 \log(SNR). 
\end{equation}

\subsection{Aperture correction}
\label{sec:psf}

Most of the sources detected in the FLS field at 24 $\mu$m are distant
and, due to the low spatial resolution of the instrument, can be considered
point-like. Therefore, a good estimate of the total flux can be obtained
by studying the aperture correction with a bright star.

A suitable bright star exists in the field. Because of its brightness,
latencies are visible in the combined image. To eliminate the
latencies from the mosaicked image, we subtracted from every BCD
a fraction of the four BCDs taken before. We used for this task 
the empirical factors 0.007,0.002,0.002, and 0.001.
The effect of the correction can be seen in Figure~\ref{fig:latency}.

For comparison, a synthetic star has been obtained using the same BCDs
around the real star with a theoretical PSF made with STinyTim (SED of
a blackbody at 6000 K). This image has been mosaicked using
the same reference frame as the real image as well as masking the same
pixels as in the real case.

\begin{figure}[t!]
\includegraphics[angle=0,width=8cm,clip=true,angle=0]{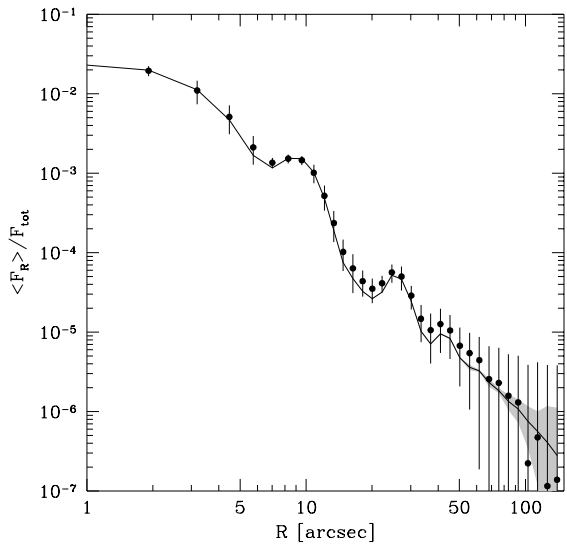}
\caption{Azimuthally-averaged profile normalized to the total flux of
the source.  Line and dots correspond to the synthetic and real
source, respectively.  The first three Airy rings are clearly visible.
Error bars correspond to the 1-$\sigma$ dispersion of fluxes inside
the annulus.  The shaded region corresponds to the same 1-$\sigma$
dispersion for the synthetic star.
\label{fig:psfprofile}
}
\end{figure}

After subtracting any point sources detected around the star, we
subtracted an average value of 0.007 MJy/sr which comes from residual
background due to contribution of faint sources and has been estimated
in regions close to the star.  We computed then the flux in circular
annuli around the real and synthetic star.  The azimuthally-averaged
profile is well reproduced by the model in the region including the
first Airy ring (Figure~\ref{fig:psfprofile}). Then, the real star
appears slightly brighter than the synthetic one between 20 and 50
arcsec, probably because of the contribution of a few faint sources
which lie on the Airy rings and we were not able to detect and
subtract.

For the radius that we are using in our extraction (11 pixels corresponding
to 14.02 arcsec) the correction is 1.160 with the synthetic star
and 1.169 with the real one. 
In Table~\ref{tbl:apcorr} we report 
the aperture corrections computed for several apertures with the
synthetic and real stars. The maximal aperture we considered in our
measurements contained 99.45\% of the total flux of the synthetic star.
In our computation we assumed the same value also in the case of the real star.

We chose to use the theoretical factor (1.160) since the agreement
with the real case is very good for radii less than 30 arcsec.  The
difference is less than 0.8\% and part of it comes probably from
contamination of faint sources which we are not able to subtract.  To
overcome this limitation, one needs to stack several bright stars
which can be found reducing many wide-field observations as, for instance, 
those in the SWIRE fields (Lonsdale et al. 2003).

We remind that, in the case of source extraction, we were able to
follow such a procedure. In fact, since in this case we consider only
the part inside the first Airy ring, we were able to build an
empirical PRF stacking several bright isolated sources to obtain a PRF
free of contamination by surrounding faint sources.

\begin{deluxetable}{ccccc}
\tablecaption{Aperture correction.\label{tbl:apcorr}}
\tablewidth{0pt}
\tablehead{
\colhead{Radius} & \colhead{Theoretical}   & \colhead{Empirical}\\
\colhead{(arcsec)} &\colhead{correction} &\colhead{correction}\\
}
\startdata
 5.10 &1.912 (52.3\%)& 1.986 (50.4\%)\\
 6.38 &1.723 (58.1\%)& 1.773 (56.4\%)\\
 7.65 &1.590 (62.9\%)& 1.610 (62.1\%)\\
 8.93 &1.467 (68.2\%)& 1.481 (67.5\%)\\
10.20 &1.323 (75.6\%)& 1.339 (74.7\%)\\
11.47 &1.224 (81.7\%)& 1.241 (80.6\%)\\
12.75 &1.182 (84.6\%)& 1.194 (83.8\%)\\
14.02 &1.160 (86.2\%)& 1.169 (85.5\%)\\
15.30 &1.153 (86.7\%)& 1.158 (86.3\%)\\
16.57 &1.147 (87.2\%)& 1.150 (86.9\%)\\
17.85 &1.142 (87.6\%)& 1.144 (87.4\%)\\
19.12 &1.139 (87.8\%)& 1.139 (87.8\%)\\
\enddata
\end{deluxetable}

\begin{figure*}[t!]
\hbox{
\includegraphics[angle=0,width=8cm,clip=true,angle=0]{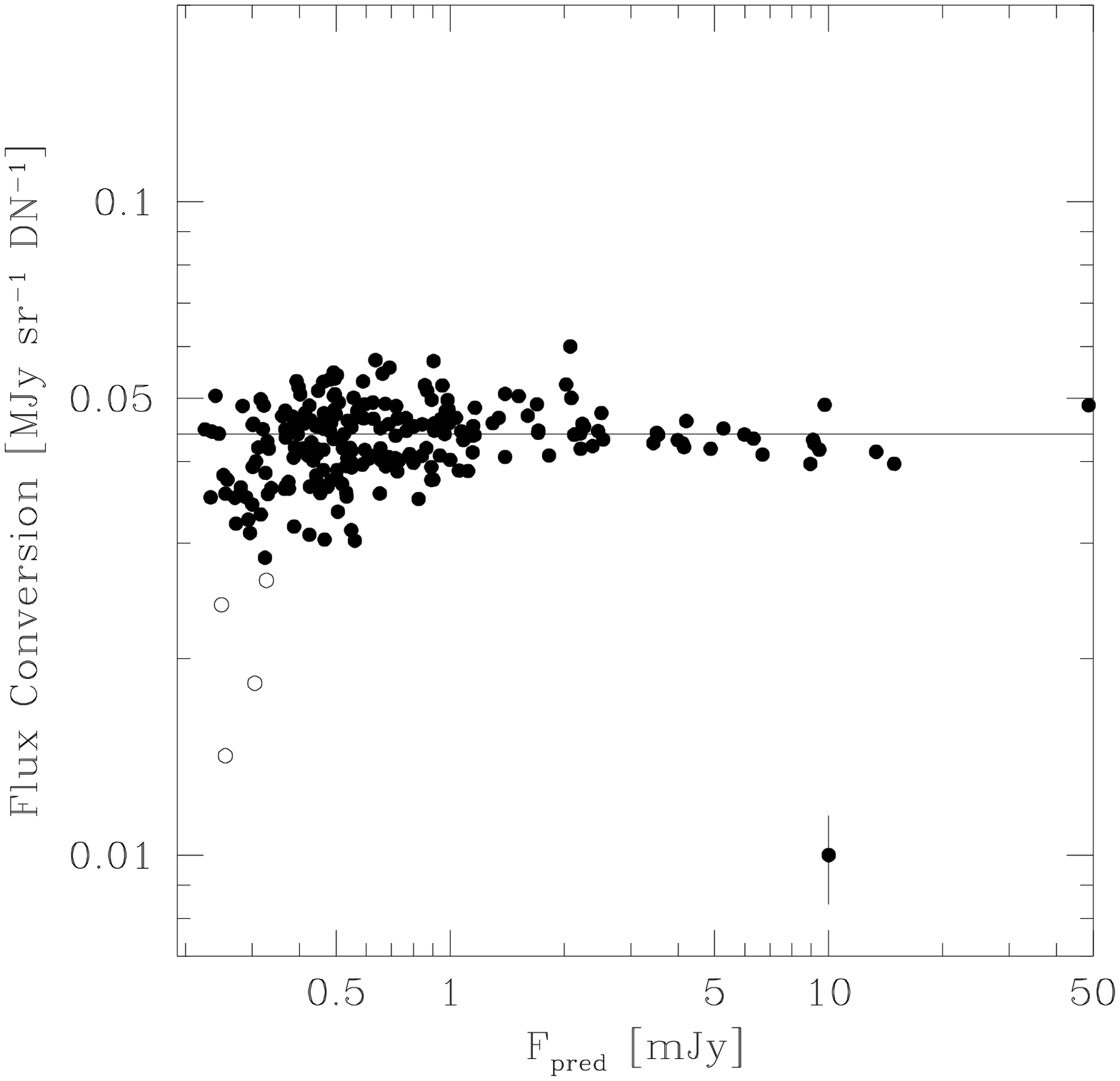}
\includegraphics[angle=0,width=8cm,clip=true,angle=0]{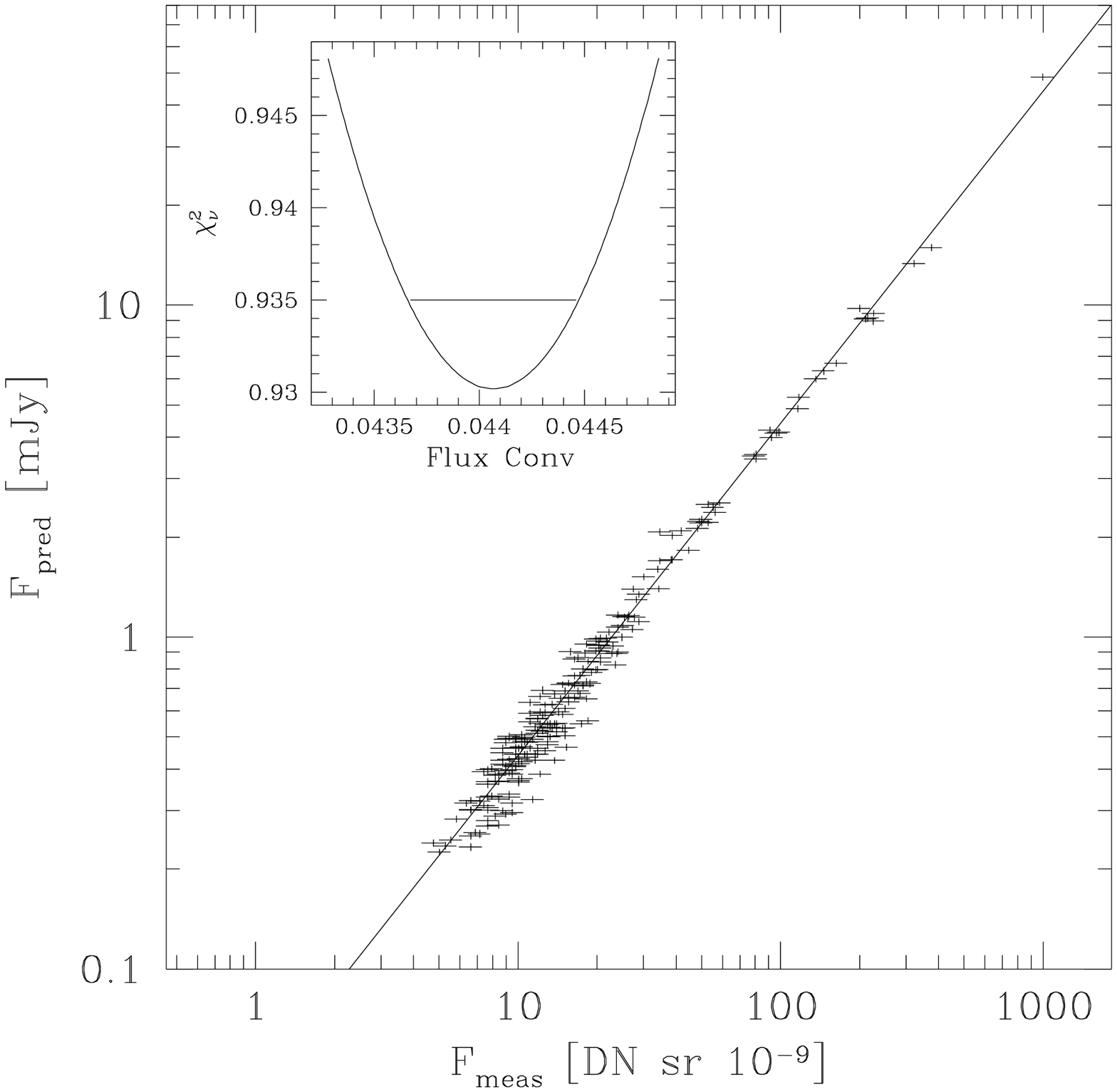}
}
\caption{ {\it Left}: calibration factor
versus predicted flux for 24$\mu$m stars in the FLS with SDSS, 2MASS
and IRAC counterparts. The point in the bottom-right corner indicates
the typical errorbar. The empty circles are stars with possible excess
of infrared emission not considered in our analysis.
{\it Right}: relationship between predicted
and measured values of the flux for the 224 stars marked with full
circles on the left plot.  The value found is less than 0.5\% higher than
the official 24$\mu$m Spitzer calibration value (see
horizontal line on the left plot). One-sigma errorbars are reported on
the plots. The inset shows the reduced $\chi^2$ as a function of the calibration
value. The horizontal line corresponds to the 1$\sigma$ confidence interval.}
\label{fig:calibration}
\end{figure*}

\subsection{Absolute Calibration}

Spitzer calibration is based on routine observations of stars during
every MIPS campaign and a set of stars with 24$\mu$m fluxes between
0.02 and 2 Jy, observed at least one or more times since the
commencement of MIPS observations.  Based on the routine star
observations, the calibration is stable from campaign to campaign
within approximately 3\%.  The uncertainty of the flux conversion
factor ( 0.04391 MJy~sr$^{-1}$~DN$^{-1}$) is less than 5\%, according
to the MIPS Data Handbook~\footnote{ssc.spitzer.caltech.edu/mips/dh}.

There are many reasons to check the calibration factor.  Calibration
stars are observed in photometry mode and the FLS field was observed
with medium-scan mode.  The scan mode, which compensates for the
motion of the telescope with the movement of the cryogenic scan
mirror, produces slightly broader PSFs than the photometric mode.
Moreover, the technique used to reduce the calibration data is not
identical to that used in this paper.  Finally, although the
calibration is remarkably stable, every campaign has a slightly
different calibration factor.

\begin{deluxetable}{cccccccccccccccc}
\setlength{\tabcolsep}{3pt}
\tablecolumns{16}
\tabletypesize{\scriptsize}
\tablecaption{Stars used for absolute calibration (first five entries).\label{tbl:stars}}
\tablewidth{0pt}
\tablehead{
\colhead{RA} & \multicolumn{5}{c}{SDSS}& \multicolumn{3}{c}{2MASS}&\multicolumn{3}{c}{IRAC}&
\multicolumn{2}{c}{$F24$}&\multicolumn{2}{c}{$T_{eff}$[K]}\\
\colhead{Dec}& \colhead{u'}& \colhead{g'}& \colhead{r'}& \colhead{i'}& \colhead{z'}&
\colhead{J}& \colhead{H}& \colhead{Ks}& \colhead{3.5$\mu$m}& \colhead{4.5$\mu$m}& \colhead{5.8$\mu$m}&
\colhead{$Meas$}&\colhead{$Pred$}&\colhead{[Fe/H]}&\colhead{$\log g$}\\
\colhead{(J2000)}&\colhead{[mag]}&\colhead{[mag]}&\colhead{[mag]}&\colhead{[mag]}&\colhead{[mag]}
&\colhead{[mag]}&\colhead{[mag]}&\colhead{[mag]}&\colhead{[mJy]}&\colhead{[mJy]}&\colhead{[mJy]}
&\colhead{[mJy]}&\colhead{[mJy]}&\colhead{[Sun]}&\colhead{[cm s$^-2$]}
}
\startdata
17:09:58.292& 15.84& 14.93*&14.89*&12.42 &13.00*&11.10 &10.55 &10.45 & 19.4 & 11.5  & 7.3 & 0.65 & 0.53&\multicolumn{2}{c}{4500}\\
+58:38:56.00&  0.06&  0.05 & 0.04 & 0.02 & 0.02 & 0.03 & 0.03 & 0.02 &  1.9 &  1.1  & 0.7 & 0.07 & 0.0 & 1.50 &-1.50\\\\
17:10:31.493& 14.75& 11.37 &10.88 &10.74 &12.96*& 9.81 & 9.51 & 9.45 & 48.5 & 31.1  &18.4 & 1.26 & 1.34&\multicolumn{2}{c}{5250}\\
+59:26:03.72&  0.06&  0.04 & 0.03 & 0.02 & 0.03 & 0.03 & 0.03 & 0.02 &  4.8 &  3.1  & 1.8 & 0.13 & 0.0 & 0.00 & 1.00\\\\
17:11:11.095& 15.74& 15.23*&12.81 &12.53 &12.99*&11.30 &10.83 &10.73 & 15.0 &  8.8  & 5.4 & 0.42 & 0.40&\multicolumn{2}{c}{4750}\\
+59:50:56.45&  0.06&  0.05 & 0.03 & 0.02 & 0.02 & 0.03 & 0.03 & 0.02 &  1.5 &  0.9  & 0.5 & 0.04 & 0.0 & 1.00 &-1.00\\\\
17:11:18.604& 16.11& 15.16*&12.87 &14.63*&12.99*&11.11 &10.47 &10.32 & 20.5 & 12.8  & 8.7 & 0.48 & 0.63&\multicolumn{2}{c}{4250}\\
+59:59:31.04&  0.06&  0.05 & 0.03 & 0.04 & 0.02 & 0.03 & 0.03 & 0.02 &  2.0 &  1.3  & 0.9 & 0.05 & 0.0 & 4.50 &-1.50\\\\
17:10:51.481& 15.23& 16.26*&10.01 &12.70*& 9.99*& 8.36 & 7.81 & 7.67 &164.2 &122.9  &93.1 & 6.40 & 6.33&\multicolumn{2}{c}{4500}\\
+58:46:07.97&  0.06&  0.07 & 0.03 & 0.03 & 0.02 & 0.03 & 0.03 & 0.02 & 16.4 & 12.3  & 9.3 & 0.64 & 0.2 & 1.00 & 1.00
\enddata
\tablenotetext{a}{Asterisks indicate that points were considered deviant in the fit.}
\end{deluxetable}

To evaluate the importance of these effects, we used a set of 279
stars in the FLS field which have Sloan (in the five Sloan bands),
2MASS (in J,H, and Ks) and IRAC counterparts (at 3.5$\mu$m, 4.5$\mu$m
and 5.8$\mu$m) within 2 arcseconds.
The main
advantage of the FLS field for calibration purposes is that the
Galactic extinction is very low (E(B-V) $\sim$ 0.022 mag).  Although
the extinction is generally lower for most of the stars, we applied a
general correction to the visible and near-infrared fluxes using the
Galactic extinction.  Optical, near-infrared and infrared fluxes are
used to fit star SEDs with a grid of Basel2.2 models (Lejeune et al.,
1998).  We do not use the 8$\mu$m IRAC fluxes for the fit since the models
do not take into account the SiO absorption feature at this
wavelength.  On the basis of the SED fits, we discarded from the
original sample 13 probable AGNs and 35 stars with infrared excess
in the IRAS and MIPS bands.  We considered only stars with
predicted flux greater than 0.2 mJy for computing the calibration
factor since for fainter stars the photometry is not reliable
enough. Furthermore, we discarded from the sample a few stars with
infrared excess at 24$\mu$m (see empty circles in Figure~\ref{fig:calibration}).
On the basis of the 224 stars considered, the flux conversion factor
is $0.0441 \pm 0.0004$\%, less than 0.5\% higher than that reported in
the MIPS Data Handbook.  To compute this value we fitted a line
passing through the origin taking into account the errors of the two
variables (see Figure~\ref{fig:calibration}). To do this, we minimized
the $\chi^2$ defined as:
\begin{equation}
\chi^2 = \sum_{i=1,N} \frac{(y_i-k\, x_i)^2}{\sigma_{yi}^2+k^2\,\sigma_{xi}^2},
\end{equation}
where $x_i$ and $y_i$ are the measured and predicted fluxes,
respectively, and $\sigma_{xi}$, $\sigma_{yi}$ the respective errors
(see inset in Figure~\ref{fig:calibration} and Press et al., 1992 for 
the $\chi^2$ formula).  The data used for the fits as well as
predicted fluxes at 24$\mu$m and the features of the best fit SEDs are
reported in Table~\ref{tbl:stars} (the entire list is available in the
electronic version of the Journal).
 Since the
Spitzer calibration is done with well known stars and our value is
well inside the error bar of the official calibration, we retained
the official flux conversion value for computing the fluxes in our catalogs.

\subsection{Color Correction}

\begin{figure}[t!]
\includegraphics[angle=0,width=8cm,clip=true,angle=0]{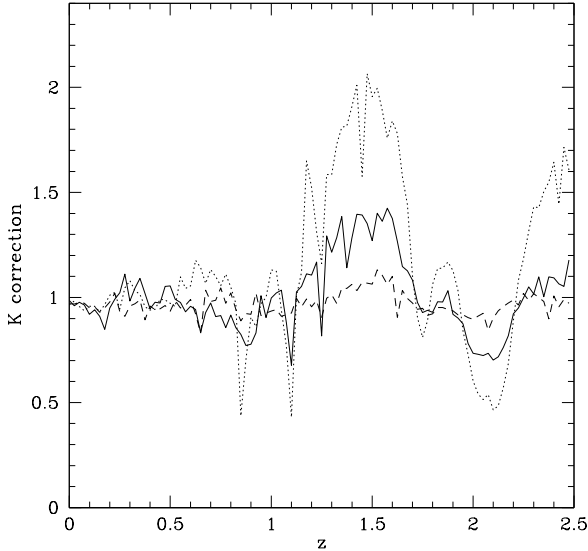}
\caption{Color correction as function of the redshift for three different
template galaxies: the starburst M82 (dotted line) and the Seyfert galaxies
 NGC1068 (dashed line) and Circinus (solid line). Color correction is
expected to be relevant only for sources beyond $z=1.1$.
\label{fig:colcorr}
}
\end{figure}

Spitzer 24$\mu$m calibrations are done with stars and the official calibration
is referred to a blackbody with an effective temperature of 10000K
(see MIPS Data Handbook). When observing extragalactic objects,
the slope of their SEDs is expected to be different from those of
stars.
According to the MIPS Data Handbook, the color correction is:
\begin{equation}
K = \frac
{\int 
\frac
{F_{\lambda}}
{F_{\lambda_0}}
\lambda R_{\lambda} d\lambda}
{
\int (\frac{\lambda_0}{\lambda})^5 
\frac
{e^{\frac{hc}{\lambda_0 k T_0}}-1}
{e^{\frac{hc}{\lambda k T_0}}-1} 
\lambda R_{\lambda} d\lambda }, 
\end{equation}
where $\lambda_0 = 23.675 \mu$m is the effective wavelength of the
MIPS 24$\mu$m band, $T_0 =10000$K and $R_{\lambda}$ is the response
function.  In Figure~\ref{fig:colcorr}, we have computed this term in
the case of three template galaxies in the range of redshift 0-2.5.
The color term is very close to 1 at least up to $z=1.1$.  At higher
redshifts, the PAH-emission and Si-absorption features fall into the filter and
huge variations are expected especially in the case of starburst
galaxies.

Since there is no systematic correction independent of the redshift,
we did not apply any color correction to the value of the fluxes
measured in our catalog.

\section{Cirrus Foreground}
\label{sct:cirri}

The careful method used to subtract zodiacal light and transient
effects allowed us to study the foreground variations at 24$\mu$m due
to molecular cloud emission.  In order to get rid of source emissions,
we subtracted the point sources extracted at the 3$\sigma$ level from
the final mosaic image using the PRF estimated from the data and the
fluxes computed with MOPEX.  For the extended sources, it is not
possible to perform such a subtraction.  Therefore, we masked the
regions containing the extended sources and substitute them with
interpolated values.  The residual image and the list of pixels masked
are part of the data released with the paper (see
section~\ref{sct:dataprod}).  To enhance the structures present in the
residual image, we degraded the resolution of the original image to a
pixel scale of 19 arcsec. The value of a pixel corresponds to the
average of 15$\times$15 pixels in the original image which greatly reduce
the noise.

\begin{figure*}[t!]
\hbox{
\includegraphics[angle=0,width=8cm,clip=true,angle=0]{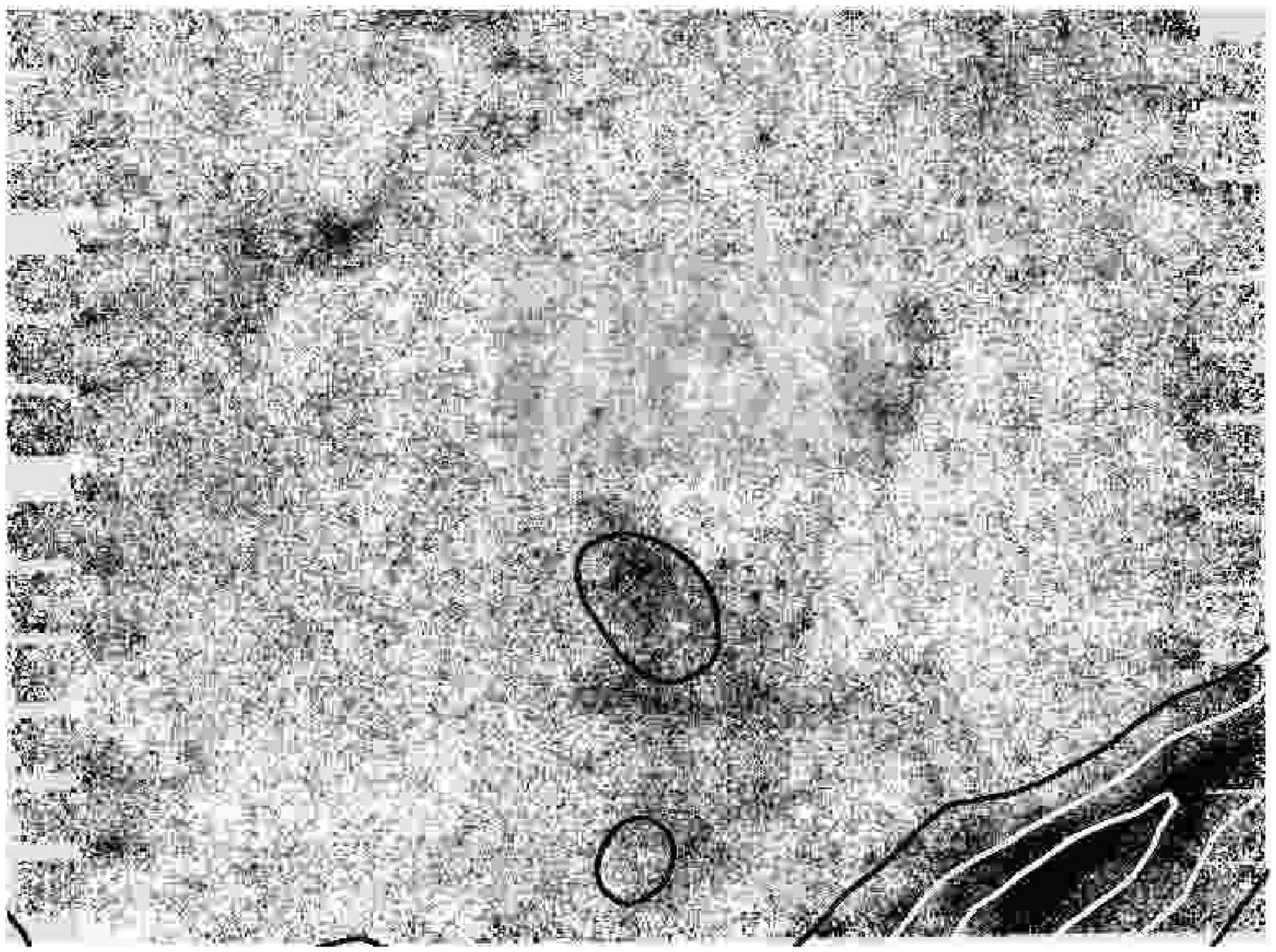}
\includegraphics[angle=0,width=8cm,clip=true,angle=0]{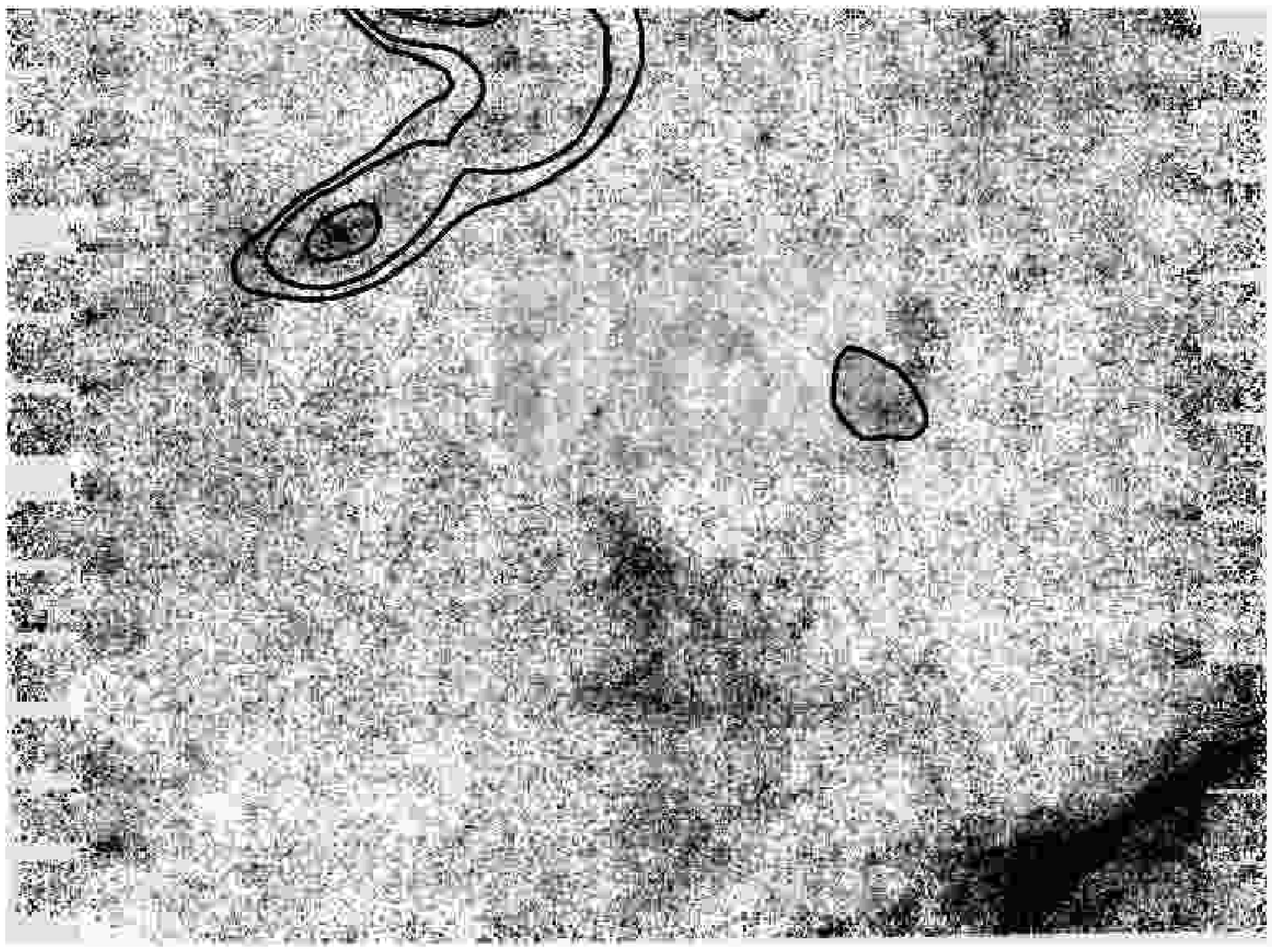}
}
\caption{Cirrus foreground at 24 $\mu$m compared with the HI maps of
Lockman \& Condon (2005). North is up and East is left. {\em Left:}
low velocity clouds ( -4.7$< V <$ 0.7 km/s) with $N_{HI}$ contours
corresponding to 0.16, 0.2 and 0.3 $10^{20} cm^{-2}$. {\em Right: }
intermediate velocity clouds (-35 $< V <$ -45 km/s) with $N_{HI}$
contours corresponding to 0.87, 1.2 and 1.5 $10^{20} cm^{-2}$.
\label{fig:cirri}
}
\end{figure*}

Several structures are clearly visible in Figure~\ref{fig:cirri}.
When compared to the neutral Hydrogen emission maps of Lockman \&
Condon (2005), the main structures correspond to low-velocity
clouds. In particular, the arc is a cloud with velocity $V_{LSR} = -2$ km/s.
Also a few intermediate-velocity structures are visible.  The main
ones are classified under the names IVC1 and IVC2 by Lockman \& Condon
(2005) and have velocities $V_{LSR} = -41$ km/s.

\section{Radio and Optical Associations}

We searched for 24$\mu$m source counterparts in the VLA 20cm and in
the KPNO-R FLS catalogs (see Condon et al. 2003, Fadda et al. 2004).
The source density of the radio catalog is less than half that of the
24$\mu$m sources (0.44 and 0.14 in the case of the main and
verification fields, respectively). Therefore it is rare to have two
or more possible counterparts for a 24$\mu$m source. Assuming a Poissonian
distribution for the VLA sources, the probability
of having  more than one counterpart within a radius of 4 arcsec is
0.4\%. This is
obviously not the case for the optical catalog which contains almost
34 times the number of sources as the 24$\mu$m catalog.  In this case
we have used a probabilistic approach to evaluate the reliability of
the associations with 24$\mu$m point sources.  Associations of
extended sources were treated separately by visually associating
infrared sources with radio and optical counterparts.  The fluxes and
magnitudes of the counterparts are reported in the detection
Table~\ref{tbl:extsrc}.

\subsection{Radio Counterparts}

82\% of the 24$\mu$m field has been surveyed at 20cm with the VLA
(Condon et al. 2003) reaching a limiting flux of 0.1 mJy (see
Figure~\ref{fig:surveys}). We performed a cross-correlation between
our 24$\mu$m point source catalog and the VLA catalog. For this
comparison, we used a catalog of VLA sources which is slightly
deeper than the 2003 version. 4904 VLA sources lie in our field.  We
found 2415 reliable associations with a positional difference of less
than 4 arcsec which roughly corresponds to three times the combined error
of radio and 24$\mu$m detections.
This means that  approximately 16\% of our sample of 24$\mu$m
sources have 20cm counterparts. There are no ambiguities in these
associations, i.e.~there is a unique counterpart found inside the chosen
radius. In three cases the 24$\mu$m point sources have
extended radio sources as counterparts.  We matched in the same way
 the extended source catalog, finding counterparts for 53\% of the
94 sources inside the VLA field.  The radio-infrared associations are given
in the point source catalog (see section~\ref{sct:dataprod}).
For each associated radio source, we report its position, the 20cm
flux and the distance between the radio and 24$\mu$m positions.

82\% of the radio associations also have a reliable optical counterpart.
This ratio is very similar to that of the infrared sources with optical
counterparts (see section~\ref{sct:optcparts}) and therefore it is essentially limited by
the depth of the optical catalog used.

\subsection{Optical Counterparts}
\label{sct:optcparts}

The optical catalog we used for matching 24$\mu$m sources is derived from the
R-band KPNO observations of Fadda et al. (2004). Since objects
brighter than R=18 are typically saturated in the KPNO data, we used
the Sloan observations (Hogg et al., in prep.) to complete the bright
magnitude range.  To transform Sloan magnitudes into R magnitudes, we
used the formula derived by Fadda et al. (2004) through calibration
stars using the r' and i' magnitudes.  The Sloan data homogeneously
cover the entire 24$\mu$m FLS field down to an R magnitude of 23.  The
KPNO data cover most of the field with the exception of three corners
(99.5\% of the total area) with slightly variable depth.  The typical
limiting R magnitude is 25.5.  We considered only the part of the
field covered by the KPNO data when searching for counterparts.  To
obtain a unique catalog from the various KPNO images, we first
adjusted the astrometry of every image to that of the SDSS using SDSS
counterparts of the KPNO sources. We then  merged the catalogs of
every image by completing them in the bright range with SDSS sources.

\begin{figure}[t!]
\includegraphics[angle=0,width=8cm,clip=true,angle=0]{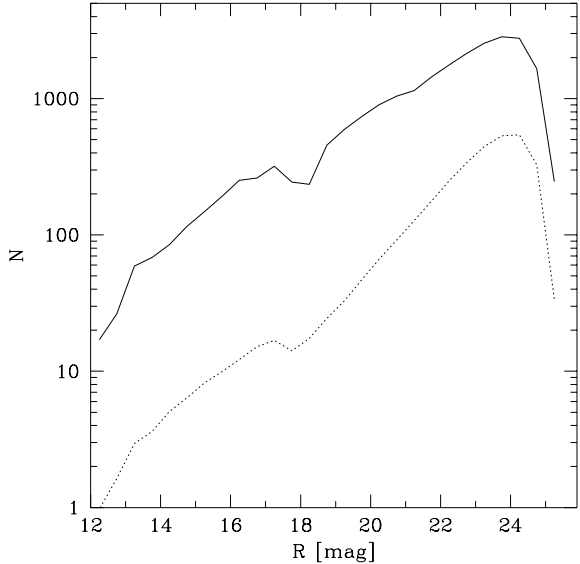}
\caption{ Magnitude distribution of possible counterparts (solid line) and
background objects (dotted line) within a 2 arcsec radius from each 24$\mu$m
source. The solid curve has been normalized according to equation~(\ref{eq:real})
and used in the Likelihood Ratio definition.
\label{fig:real_bgr}
}
\end{figure}

To compute the reliability of the optical associations we used the 
likelihood ratio technique described by Sutherland \&  Saunders (1992)
following the Ciliegi et al. (2003) practical implementation.
The likelihood ratio is defined as the ratio between the probability that
an optical source is the real counterpart and the probability that the 
association is actually with a background object.
Assuming that magnitudes and positions are not correlated, we can 
factorize the probability that a counterpart lies at a radius $r$ from the
source and have a magnitude $m$:
\begin{equation}
p(r,m) = f(r) \, 2\pi r dr \, g(m) \, dm.
\end{equation}
Therefore, the likelihood ratio is:
\begin{equation}
L = \frac{f(r) g(m)}{b(m)},
\end{equation}
with $b$ the surface density of background objects at magnitude $m$.
The reliability of an association with the $i$-th candidate  is defined as:
\begin{equation}
R_i = \frac{L_i}{\sum_j L_j + (1-G)},
\end{equation}
where the sum is done over the set of all possible candidates
and $G$ is the probability that the candidate is brighter than the 
magnitude limit of the catalog ($ G = \int^{m_{lim}} g(m) dm $).

To estimate $g(m)$ we computed the overdensity of optical sources with 
respect to the background around each 24$\mu$m source.
In order to maximize the number of real counterparts we chose a radius
approximately twice the 1-$\sigma$ error in position ($r_0 = $1.6 arcsec).
For each magnitude bin (0.5 mag between 16 and 25.5), we have an 
overdensity of possible counterparts with respect to the background:
\begin{equation}
o(m) = \sum_{i} N_{c,i} - N \, \pi r_0^2 \, b(m),
\end{equation}
with $N$, the number of 24$\mu$m point sources and $N_{c,i}$,
 the number of possible counterparts
 for the source $i$ inside the circle of radius $r_0$ (see Figure~\ref{fig:real_bgr}).
To obtain the $g$ function, we normalize the overdensity and correct
for the limiting magnitude of our catalog:
\begin{equation}
g(m) = \frac{o(m)}{\sum_i o(m_i)} \, G,
\label{eq:real}
\end{equation}
with G estimated on the basis of the ratio of the number of associations
with the total number of 24$\mu$m sources.
We fixed $G=0.8$, approximately the ratio of sources with optical counterparts.
The number of counterparts, as well as their reliability, does not
change substantially for $G$ in the range $0.7-0.9$.

As the probability function for the positional error we adopted a Gaussian
distribution whose standard deviation takes into account the combination
of infrared and optical position accuracy.

We searched for possible counterparts inside a circle of 5 arcsec
(approximately 5-$\sigma$ in radial position error) and accepted as
possible counterparts the sources with likelihood ratio greater than
$L_{th} = 0.2$.  With this threshold, the optical counterparts with
only one identification (93\% of the sample) and $L > L_{th}$ have a
reliability greater than 0.5.  With this threshold we find that 13783
out of the 16823 sources detected at 24$\mu$m within the area of the KPNO images have a likely
identification, approximately 82\%. 946 sources have more than one
possible counterpart with $L > L_{th}$ , that is 6\% of the
associations. In these cases, we assumed  the object with the highest
Likelihood ratio value as the counterpart of the 24$\mu$m source.
The reliability of the association of sources with a unique counterpart is usually
very high (0.9 in 92\% of the cases).

The results of the optical identifications are reported in the point
source catalog (see section~\ref{sct:dataprod}). For each 24$\mu$m source we
report the coordinates of the most probable optical counterpart within
5 arcsec, its $R$ magnitude, the distance between the infrared and
optical sources, and the reliability of the association.

\section{Data Products}
\label{sct:dataprod}

The images and catalogs produced in the analysis of the survey data
are made available at the {\em Spitzer Science Center}
science archive~\footnote{using the {\it Leopard} archive tool or directly
at http://data.spitzer.caltech.edu/popular/fls}.

\subsection{Images}
The images which are made publicly available are in FITS format 
and have a size of 171~Mbyte. The header contains the astrometric
information and the tangential projection is used.

{\em Intensity map:}  image with pixel size of 1.275 arcsec and map units
of MJy/sr. The image corresponds to the mosaic of the original BCDs
reduced as explained in section~\ref{sct:datared}.

 {\em Coverage map:}  image with pixel size of 1.275 arcsec and map units
of number of BCDs. Since the integration time is 3.67 seconds per BCD,
the exposure map can be obtained simply by multiplying this map by
the exposure time per BCD.

 {\em Uncertainty map:}  map with a resolution of 1.275 arcsec which has been
used for the extraction of source positions. It has been computed with MOPEX
assuming a Poissonian noise plus a Gaussian component due to the readout noise
with dispersion of 40.~e$^-$, mean of 0.~e$^-$ and a gain of 5.~e$^-$/DN.

 {\em Residual map:} map with resolution of 1.275 arcsec obtained
by subtracting the point sources and masking and interpolating the regions
occupied by extended sources.

 {\em Bad Pixel mask:} mask defining the regions occupied by extended
sources which have been interpolated in the residual map.  The file is
given in the pixel-list IRAF format.

\subsection{Catalogs}

Three catalogs are given.  The list of the 123 extended sources and
their optical/radio counterparts (see Table~\ref{tbl:extsrc}) and the
list of 231 stars considered for the calibration (Table~\ref{tbl:stars}) are
offered in the complete form in the electronic version of the Journal.

The catalog of point sources including the optical and radio
associations is an ASCII Table with the format explained in
Table~\ref{tbl:cat} available in the electronic version of the
Journal.  The entries are ordered by decreasing signal-to-noise ratio
of 24$\mu$m detections.  The catalog contains 16905 sources detected
with a signal-to-noise ratio greater than 5.  The catalog reports the
position errors and flux errors estimated from the simulations. The
position error includes the effect of extraction and pointing errors.
The flux errors do not account for the systematic error due to the
calibration.

\begin{deluxetable}{lcl}
\setlength{\tabcolsep}{3pt}
\tablecolumns{3}
\tablecaption{Format of the 24$\mu$m Point Source Catalog.\label{tbl:cat}}
\tablewidth{0pt}
\tablehead{
\colhead{Column} & \colhead{Format}&\colhead{Description}
}
\startdata
   1:16 &a16 & IAU designation \\
  18:19&i2  & R.A. hours (J2000) \\
  20:21&i2  & R.A. minutes (J2000)\\ 
  22:27&f6.3& R.A. seconds (J2000) \\
  28   &a1  & Sign of declination \\
  29:30&i2  & Dec. degrees (J2000) \\
  31:32&i2  & Dec. minutes (J2000) \\
  33:37&f5.2& Dec. arcseconds (J2000) \\
  39:41&f3.1& Position error in arcseconds \\
  43:47&f5.2& 24$\mu$m flux in mJy \\
  49:52&f4.2& 24$\mu$m flux error in mJy \\
  54:58&f5.1& Signal-to-noise ratio of the detection\\
  60:61&i2  & R.A. hours (J2000) of the optical counterpart\\
  62:63&i2  & R.A. minutes (J2000) of the optical counterpart \\
  64:69&f6.3& R.A. seconds (J2000 of the optical counterpart) \\
  70   &a1  & Sign of declination of the optical counterpart \\
  71:72&i2  & Dec. degrees (J2000) of the optical counterpart \\
  73:74&i2  & Dec. minutes (J2000) of the optical counterpart \\
  75:79&f5.2& Dec. arcseconds (J2000) of the optical counterpart \\
  81:85&f5.2& Magnitude R of the optical counterpart \\
  87:89&f3.1& Distance between infrared source and optical counterpart in arcsec \\
  91   &i1  & Number of possible optical counterparts within 5 arcsec \\
  93:96&f4.2& Reliability of the most likely optical counterpart \\
  98&i1  & Index of coverage: 1 if out of KPNO field, 0 if inside the KPNO field \\
100:101&i2  & R.A. hours (J2000) of the radio counterpart\\
102:103&i2  & R.A. minutes (J2000) of the radio counterpart \\
104:109&f6.3& R.A. seconds (J2000 of the radio counterpart) \\
110    &a1  & Sign of declination of the radio counterpart \\
111:112&i2  & Dec. degrees (J2000) of the radio counterpart \\
113:114&i2  & Dec. minutes (J2000) of the radio counterpart \\
115:119&f5.2& Dec. arcseconds (J2000) of the radio counterpart \\
121:126&f6.2& 20~cm flux in mJy of the radio counterpart \\
128:130&f3.1& Distance between infrared source and radio counterpart in arcsec\\
132    &i1  & Index of coverage: 1 if out of VLA field, 0 if inside the VLA field
\enddata
\end{deluxetable}

\section{Summary}

A composite image of the 24$\mu$m data in the First Look Survey region
($17^h18^m +59^o30'$) was obtained with SSC software packages and
customized routines developed to further refine the reduction.  In
particular, self-calibration flats, illumination correction and
long-term transient correction were applied to the data to get rid of
artificial variations of the background.  A catalog of approximately
17000 sources was extracted at the 5-$\sigma$ level using a PSF
fitting technique (using the MOPEX package). The PSF was estimated
from bright point sources in the field. For 123 extended sources
detected in the field which cover less than 0.3\% of the total area of
the survey, aperture photometry was used.  Aperture corrections and
the absolute calibration are estimated using stars in the field. In
particular, using a set of 224 stars with IRAC fluxes and optical and
near-IR magnitudes we found an excellent agreement with the
calibration factor used in the SSC pipeline.  Completeness,
reliability and errors in astrometry and photometry are estimated
through simulations. The 5$\sigma$ cut of the catalog guarantees
high reliability of the extracted sources.  The main and the verification
surveys are 50\% complete at 0.3 and 0.15~mJy, respectively. The
astrometric and photometric errors are functions of the SNR of the
source. Astrometric errors vary from 0.3 up to 1.1 arcsec for
sources with decreasing SNR.  Photometric errors (excluding the
uncertainty in the calibration) reach 15\% at the 5$\sigma$ level
and are approximately 5\% for sources with SNR~$\sim 20$.

The 24$\mu$m catalog has been cross-correlated with the KPNO-R
and VLA-20cm catalogs of the field (Fadda et al. 2004, Condon et al.
2003). More than 80\% of the 24$\mu$m sources have a reliable counterpart
down to the limiting depth of $R=25.5$. 6\% of these sources have 
multiple possible counterparts within 5 arcseconds. 
Approximately 16\% of the 24$\mu$m sources have a 20cm counterpart (down
to 0.1 mJy) and $\sim$80\% of these associations have a reliable optical
counterpart. 

Finally, subtracting the detected point sources from the image down to
the 3$\sigma$ level and masking and interpolating the regions occupied
by extended sources, we obtained a residual map which clearly shows
the emission from extended Galactic cirri. By comparison with the HI
data of Lockman \& Condon (2005) we detect several clouds at low and
intermediate velocities.

\acknowledgments

D.F. is grateful to Alberto Noriega-Crespo for enlightening
discussions about MIPS data processing issues. We acknowledge also the
anonymous referee for his useful comments and suggestions.


\end{document}